\documentclass[useAMS,usenatbib]{mnras}
\usepackage{epsfig}
\usepackage{color} 
\usepackage{natbib}
\usepackage{graphicx}
\usepackage{longtable}
\usepackage{booktabs}
\usepackage{array}
\def\ew{$W_{2796}$}
\def\hi{H~{\sc i}~} 
\def\nhi{$N${\sc (H~i)}}

\def\ls{L$^{\star}$~}
\def\lsoii{L$^{\star}_{[\rm O~{\sc II}]}$}
\def\lsoiii{L$^{\star}_{[\rm O~{\sc III}]}$}
\def\ssfr{$\Sigma$$_{\rm sfr}$}
\def\sloii{$\Sigma$$_{\rm [O~II]}$}
\def\loii{L$_{[\rm O~II]}$}
\def\mdyn{$M_{\rm dyn}$}
\def\sigoii{$\sigma_{[\rm O~II]}$}
\def\loiii{L$_{[\rm O~III]}$}
\def\mgi{Mg~{\sc i}} 
\def\mgii{Mg~{\sc ii}~}

\def\mgiiab{Mg~{\sc ii}$\lambda\lambda$2796,2803~} 
\def\feii{Fe~{\sc ii}~}

\def\oiiiab{[O~{\sc iii}]$\lambda\lambda$4959,5007} 
 
\def\oiiib{[O~{\sc iii}]$\lambda$5007~}

\def\oiii{[O~{\sc iii}]} 
\def\oiiab{[O~{\sc ii}]$\lambda\lambda$3727,3729}  
  
\def\oii{[O~{\sc ii}]}
\def\ohb{[O~{\sc iii}]/H$\beta$}
\def\o3o2{[O~{\sc iii}]/[O~{\sc ii}]}
\def\zabs{$z_{\rm abs}$}

\def\lya{Ly$\alpha$~}

\def\hbeta{H$\beta$~}
\def\hbetalam{H$\beta\lambda$4862~}

\def\kms{km\ s$^{-1}$} 
\def\aj{{AJ}}%
\def\apj{{ApJ}}%
\def\apjl{{ApJ}}%
\def\apjs{{ApJS}}%
\def\aap{{A\&A}}%
\def\mnras{{MNRAS}}%
\def\pasp{{PASP}}%
\def\pasj{{PASJ}}%

\title[Nature of \mgii absorber galaxies]{\oii\ nebular emission from
  \mgii absorbers: Star formation associated with the absorbing gas}
\author[Joshi, R. et al.]{Ravi Joshi$^{1}$\thanks{E-mail: rjoshi@iucaa.in(RJ)}, Raghunathan Srianand $^{1}$, Patrick  Petitjean$^{2}$  and  Pasquier Noterdaeme$^{2}$  \\
$^{1}$Inter-University Centre for Astronomy and Astrophysics, Post Bag 4, Ganeshkhind, Pune 411007, India \\
$^{2}$UPMC-CNRS, UMR7095, Institut d'Astrophysique de Paris, F$-$75014 Paris, France \\
}

\begin{document}
\date{Accepted ---. Received ---; in original form ---}

\pagerange{\pageref{firstpage}--\pageref{lastpage}} \pubyear{2016}

\maketitle

\label{firstpage}
\begin{abstract}

We present nebular emission associated with 198 strong \mgii absorbers
at 0.35 $\le z \le$ 1.1 in the fibre spectra of quasars from the Sloan
Digital Sky Survey. Measured \oii\ luminosities (\loii)
are typical of sub-\ls galaxies with derived star formation rate
(uncorrected for fibre losses and dust reddening) in the range of
0.5-20 ${\rm M_\odot\ yr^{-1}}$. Typically less than $\sim 3\%$ of the
\mgii systems with rest equivalent width, \ew $\ge 2$\AA, show
\loii\ $\ge 0.3$\lsoii. The detection rate is found to increase with
increasing \ew\ and $z$. No significant correlation is found between
\ew\ and \loii\ even when we restrict the samples to narrow
$z$-ranges. A strong correlation is seen between \loii\ and $z$. While
this is expected from the luminosity evolution of galaxies, we show
finite fibre size plays a very crucial role in this correlation. The
measured nebular line ratios (like \o3o2 and \ohb) and their $z$
evolution are consistent with those of galaxies detected in deep
surveys. Based on the median stacked spectra, we infer the average
metallicity (log Z $\sim$8.3), ionization parameter (log $q$
$\sim$7.5) and stellar mass (log (M/M$_\odot$)$\sim$9.3).  The
  \mgii systems with nebular emission typically have  \ew\ $\ge
2$~\AA, \mgii doublet ratio close to 1 and W(Fe~{\sc
  ii}$\lambda$2600)/\ew $\sim 0.5$ as often seen in damped Ly$\alpha$
and 21-cm absorbers at these redshifts. This is the biggest reported
sample of \oii\ emission from \mgii absorbers at low impact parameters
ideally suited for probing various feedback processes at play in $z\le
1$ galaxies.

\end{abstract}
\begin{keywords}
galaxies: evolution – galaxies: star formation - galaxies: ISM -
quasars: absorption lines -- cosmology:observations
\end{keywords}

%\maketitle

\section{Introduction}
\label{sec:intro_mgiidndz}

Absorption line systems traced by \mgiiab doublet in the quasar
spectra provide a direct tracer of cool, $T \sim 10^4\ K$, gas within
gaseous haloes and circumgalactic medium (CGM) surrounding galaxies
over a wide range of \hi column densities, $16 \le$ log~[\nhi\ $\rm
  cm^{-2}$] $\le 22$
\citep{Bergeron1986A&A...155L...8B,Churchill2000ApJS..130...91C,
  Rao2000ApJS..130....1R,Rigby2002ApJ...565..743R}, out to projected
distances of $\sim$ 200 kpc
\citep[e.g.,][]{Kacprzak2008AJ....135..922K, Chen2010ApJ...714.1521C,
  Chen2010ApJ...724L.176C,
  Bordoloi2011ApJ...743...10B,Nielsen2013ApJ...776..115N}. The
paradigm developed thus far is that the \mgii absorbers are linked
with either of the galactic-scale outflows originating from their host
galaxies \citep{Bouche2006MNRAS.371..495B,Tremonti2007ApJ...663L..77T,
  Weiner2009ApJ...692..187W,Martin2009ApJ...703.1394M,
  Noterdaeme2010MNRAS.403..906N, Chelouche2010ApJ...722.1821C,
  Rubin2010ApJ...712..574R,Lundgren2012ApJ...760...49L,
  Bordoloi2014ApJ...794..130B}, dynamical mergers or filamentary
accretion on to galaxies \citep{Steidel2002ApJ...570..526S,
  Chen2010ApJ...714.1521C, Kacprzak2010ApJ...711..533K,
  Kacprzak2011MNRAS.416.3118K, Kacprzak2012MNRAS.427.3029K,
  Martin2012ApJ...760..127M, Rubin2012ApJ...747L..26R} and high
velocity clouds [HVCs]
\citep{Richter2012ApJ...750..165R,Herenz2013A&A...550A..87H}. Thus
\mgii absorbers may be tracing different feedback processes that
control the evolution of their host galaxies.

 In efforts to determine physical properties of galaxies associated
 with \mgii absorbers (hereinafter \mgii galaxies), the rest-frame
 equivalent width (\ew) of \mgii absorption is found to be
 anticorrelated with impact parameter ($\rho$), at $\sim$ 7.9$\sigma$
 level
 \citep[][]{Bergeron1991A&A...243..344B,Lanzetta1990ApJ...357..321L,Steidel1995qal..conf..139S,Churchill2000ApJS..130...91C,Kacprzak2008AJ....135..922K,Chen2010ApJ...714.1521C,Rao2011MNRAS.416.1215R,Nielsen2013ApJ...776..115N,Kacprzak2013ApJ...777L..11K}.
 This has led to the suggestions that the \mgii absorption could
 originate from the extended halos of normal galaxies with unit
 covering factor for the absorbing gas \citep[see
   e.g.,][]{Petitjean1990A&A...231..309P,Srianand1994ApJ...428...82S}.
 Recent studies suggest that the scatter in the \ew\ versus $\rho$
 relationship could be further reduced if luminosity dependent radial
 extent of the gas is also taken into account
 \citep{Chen2010ApJ...714.1521C,Nielsen2013ApJ...776..115N}. On the
 contrary, search for \mgii absorption from galaxies close to the QSO
 line-of-sight suggested that the gas covering factor is less than
 unity \citep{Tripp2005ApJ...619..714T,Barton2009AJ....138.1817B}. The
 \mgii gas distribution may be patchy around galaxies with the gas
 covering factor decreasing with the increasing impact parameter
 \citep[][]{Chen2010ApJ...714.1521C,Nielsen2013ApJ...776..115N}. While
 luminous galaxies are frequently identified at the redshift of the
 \mgii absorption, question remains that what fraction of these
 systems are produced by undetected low luminosity galaxies at much
 smaller impact parameters \citep[see figure 10 of
 ][]{Noterdaeme2010MNRAS.403..906N}.

The fibre fed spectroscopic observations, e.g., 3 and 2 arcsec
diameter fibres employed in Sloan Digital Sky Survey Data Release 7
(SDSS$-$DR7; \citealt{Abazajian2009ApJS..182..543A}) and SDSS$-$DR12
\citep[BOSS,][]{Alam2015ApJS..219...12A}, register photons from all
the objects that happen to fall within the fibre along our
line-of-sight. This in principle allows detection of nebular emission
associated with the absorbing gas
\citep{Wild2007MNRAS.374..292W,Noterdaeme2010MNRAS.403..906N,Borthakur2010ApJ...713..131B,York2012MNRAS.423.3692Y,Straka2015MNRAS.447.3856S}.
\citet{Noterdaeme2010MNRAS.403..906N} have performed an
absorption-blind search, i.e. without a prior information on
absorption, of nebular emission features (e.g., \oii, \oiii\ and
\hbeta) and detected 46 \oiii\ emitting galaxies at $0 < z < 0.8$ on
top of the quasar spectra (hereinafter refer to as GOTOQs), out of
which $\sim$17 are \mgii absorbers at $z \ge 0.4$. The
\oiii\ luminosities of these systems are found to be less than that of
an \ls galaxy with a median luminosity of 0.2 \lsoiii\ and a typical
star formation rate (SFR) in the range of 0.2-20 $\rm
M_{\odot}\ yr^{-1}$. \citet{Straka2015MNRAS.447.3856S} have
subsequently increased the number of known GOTOQs to 103 at a median
$z$ of 0.15.

A strong correlation is found between the \oii\ luminosity surface
density (\sloii) and \ew\ in the stacked spectra
\citep{Noterdaeme2010MNRAS.403..906N,Menard2011MNRAS.417..801M}. This
relationship is also found to be evolving with redshift in the sense
that a given \ew\ seems to be associated with larger \sloii\ at
high-$z$ compared to that at low-$z$.
\citet{Menard2011MNRAS.417..801M} have suggested that the \mgii
absorbers can in principle provide a new probe of the redshift
evolution of the star formation rate density in a luminosity unbiased
manner \citep[however, see][]{Lopez2012MNRAS.419.3553L}. Increasing
the number of direct detections over a large redshift range will allow
us to address this more efficiently. With an aim to probe the nature
of \mgii absorbers we have searched for the nebular emission from
\mgii absorption systems by utilizing the unprecedented number of
these systems found in the SDSS survey \citep{Zhu2013ApJ...770..130Z}.
These direct detections will not only provide the largest sample to
study {Mg~{\sc ii}}$-$galaxy correlation at low impact parameters over a large
redshift range but also help us understand the origin of
\loii\ versus \ew\ correlation seen in the stacked spectra and its
redshift dependence so that the utility of \mgii absorbers as a new
tracer of star formation in the Universe can be examined.

This article is organized as follows. Section 2 describes our sample
selection criteria and procedure for detecting the nebular emission
lines. In Section 3, using the repeat observations of QSOs in SDSS-DR7
and SDSS-DR12 we examine the effect of fibre size on the nebular
emission line detections. In Section 4, we present the absorption line
properties of systems detected in nebular emission, the detection
probability of nebular emission in \mgii systems, and discuss the
dependence of \oii\ line luminosity on \ew\ and $z$. In this section
details of observed nebular line ratio and their redshift evolution
are also discussed. We also compare the properties derived based on
nebular emission with those based on absorption lines. The summary of
our study is presented in Section 5. Throughout, we have assumed the
flat Universe with $H_0 =$ 70~\kms\ $\rm Mpc^{-1}$, $\Omega_{\rm m} =
0.3$ and $\Omega_{\rm \Lambda} = 0.7$.

 \begin{table}
 \centering
 \begin{minipage}{120mm}
 {\scriptsize
 \caption{ Sample of \mgii absorbers searched for nebular emission.}
 \label{tab:sample}
 \begin{tabular}{@{} l l r c c c c @{}}
 \hline  \hline 
 \multicolumn{1}{c}{Catalog}   &\multicolumn{1}{c}{z}  &       \multicolumn{1}{c}{N\textcolor{blue}{$^a$}}  &  \multicolumn{1}{c}{\oii\ + [O{\sc~iii}]\textcolor{blue}{$^b$}} &  \multicolumn{1}{c}{\oii\textcolor{blue}{$^c$}}&  \multicolumn{1}{c}{\oiii\textcolor{blue}{$^d$}}\\ 
\hline
SDSS-DR7                       & $ 0.35 \le$ \zabs\ $\le 0.8$      & 11523   & 29 &33 & 6 \\ 
SDSS-DR12                      & $ 0.30 \le$ \zabs\ $\le 1.1$      & 37038   & 76 &47 & 7\\
 \hline                                                                                 
 \end{tabular} 
 }             
\\\\
{$^a$ Total number of \mgii absorbers within the desired $z$ range.}\\     
{$^b$ \oii\ is detected at $\ge 4\sigma$ and  [O{\sc~iii}] is detected at $\ge 3\sigma$}\\
{$^c$ \oii\ is detected at $\ge 4\sigma$ and  $3\sigma$ upper limits on [O{\sc~iii}].} \\                            
{$^d$ $3\sigma$ upper limits on [O{\sc~ii}]  and  [O{\sc~iii}] is detected at $\ge 3\sigma$}\\                            
 \end{minipage}
 \end{table}

 \begin{figure}
 \epsfig{figure=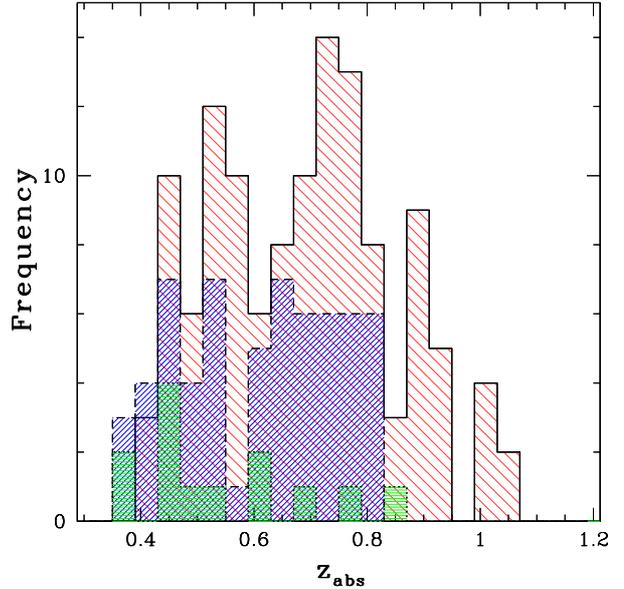,height=8.2cm,width=8.2cm}
  \caption{ The redshift distribution of \mgii absorbers detected in
    \oii\ emission in SDSS-DR7 (\emph{dashed histogram}, shaded with
    slanted lines at 45$^{\circ}$) and in SDSS-DR12 (\emph{solid
      histogram}, shaded with slanted lines at $-45^{\circ}$) spectra.
    The \emph{dotted histogram} (shaded with horizontal lines) show
    the systems where \oii\ and \oiii\ emission are detected at $<
    4\sigma$ and $\ge 3\sigma$, respectively.}
\label{fig:zabshist}
 \end{figure}

\section{Sample}
\label{lab:sample}

 In this work, we use the compilation of \mgii systems from the
 \emph{expanded-version} of JHU-SDSS Metal Absorption Line
 Catalog{\footnote{\href{http://www.pha.jhu.edu/$\sim$gz323/Site/Download\_Absorber\_Catalog\_files/fits/}
     {http://www.pha.jhu.edu/$\sim$gz323/Site/}}
   \citep[][]{Zhu2013ApJ...770..130Z}, generated from the SDSS$-$DR7
   and SDSS$-$DR12. In order to detect the nebular emission line
   features from the \mgii absorbers, first we have ensured that the
   most prominent \oii\ and \oiii\ emission lines are covered within
   the observed wavelength range of SDSS-DR7 (3800~\AA$-$9200~\AA) and
   SDSS-DR12 (3650~\AA$-$10,400~\AA) spectra. This criterion limits
   the redshift range over which \oii\ emission can be searched to
   $0.35 \le$ \zabs\ $\le 0.8$ and $0.32 \le$ \zabs\ $\le 1.0$ for
   SDSS-DR7 and SDSS-DR12 spectra respectively. This resulted in our
   primary sample of 11,000 and 37,000 \mgii absorbers with \mgii
   equivalent width ($W_{2796}$) $\ge 0.1$~\AA\ in SDSS-DR7 and DR12
   catalog respectively, where we have searched for the nebular
   emission lines.

\subsection{Search for nebular emission}
\label{lab:method}
For the redshift range of interest for the present study, the
\oiiab\ emission doublet is the strongest and most suitable nebular
line that can be detected over a wide $z$ range in a region free from
the most crowded telluric lines (e.g., O{\sc~i}$\lambda5577$,
O{\sc~i}$\lambda6300$, and OH lines). Also \oii\ is regarded as a good
indicator of the ongoing star formation rate (SFR). For each \mgii
absorber, we search for the \oiiab\ nebular emission lines at the
expected location for \zabs\ in the continuum subtracted spectrum. At
first, we model the local continuum that includes the continuum light
from both the quasar and the galaxy by a low-order (typically a third
order) polynomial fit. The significance of the detection of an
emission line feature is determined based on the signal-to-noise ratio
[$SNR$]
~\citep[see,][]{Hewett1985MNRAS.213..971H,Bolton2004MNRAS.348L..43B}
defined as:

 \begin{equation}
 SNR = \frac{ \sum_i f_i u_i/\sigma_i^2}{\sqrt{\sum_i  u_i^2/\sigma_i^2}} 
 \end{equation} 

 \noindent here, ${f_i}$ is the line flux in $i$th pixel, $\sigma_i$
 is the flux error and $u_i$ is a Gaussian kernel, normalized such
 that $\Sigma_i\ u_i = 1$, and its position and width given by the
 fitted line parameters. Any feature at the expected location of the
 \oii\ line with $SNR\ \ge 4$ is considered as a positive detection in
 this work.

 \begin{figure}
 \epsfig{figure=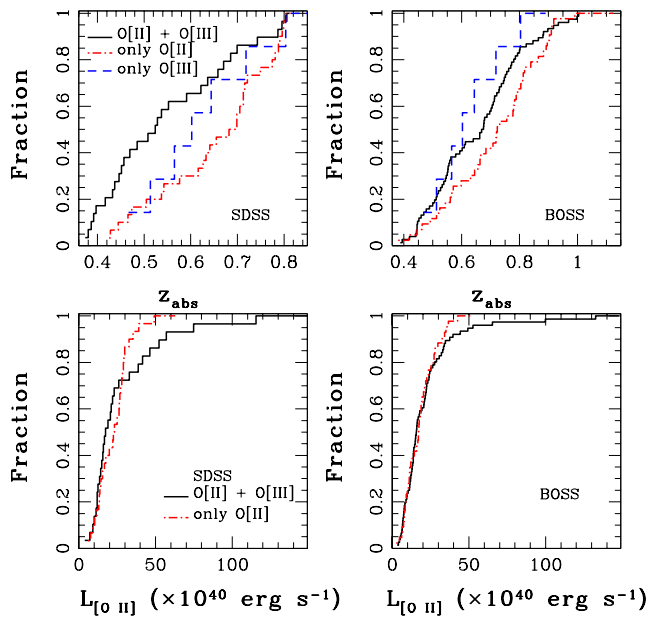,height=8.5cm,width=8.5cm}
  \caption{\emph{Top panel:} cumulative distribution of \zabs, for the
    \mgii absorbers with clear detection of both \oii\ and
    \oiii\ emission (\emph{solid}), with only \oii\ detection
    (\emph{dot dashed}) and with only \oiii\ detection (\emph{dashed})
    in SDSS-DR7 (\emph{top left}) and SDSS-DR12 (\emph{top right}).
    \emph{Bottom panel:} The same for the \oii\ luminosity.}
\label{fig:zabscomp}
 \end{figure}

\begin{figure*}
     \begin{center}
     \begin{tabular}{m{10.5cm}m{1.9cm}}

\epsfig{figure=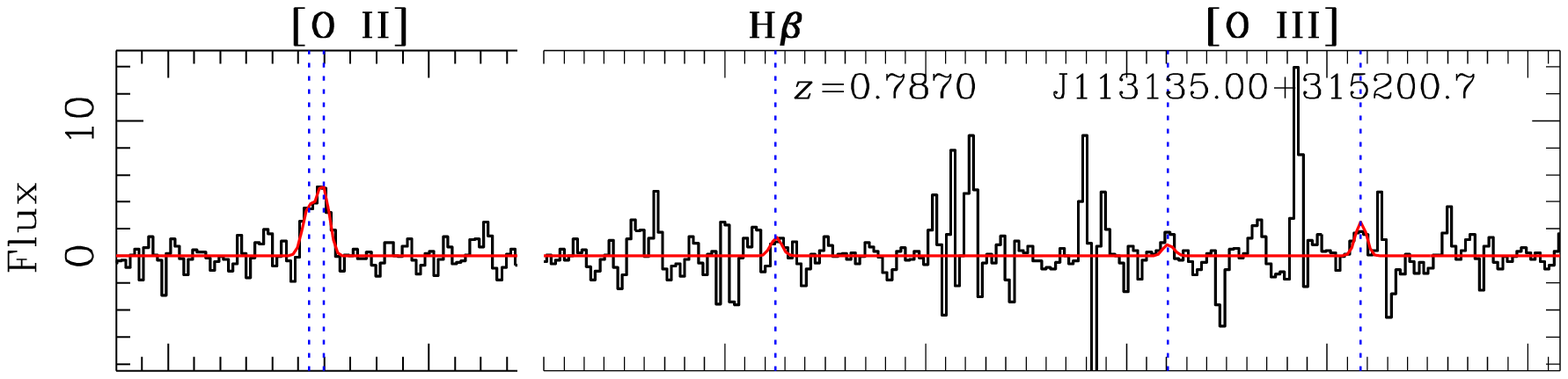,height=2.7cm,width=10.5cm,bbllx=32bp,bblly=190bp,bburx=550bp,bbury=337bp,clip=true}
& \epsfig{figure=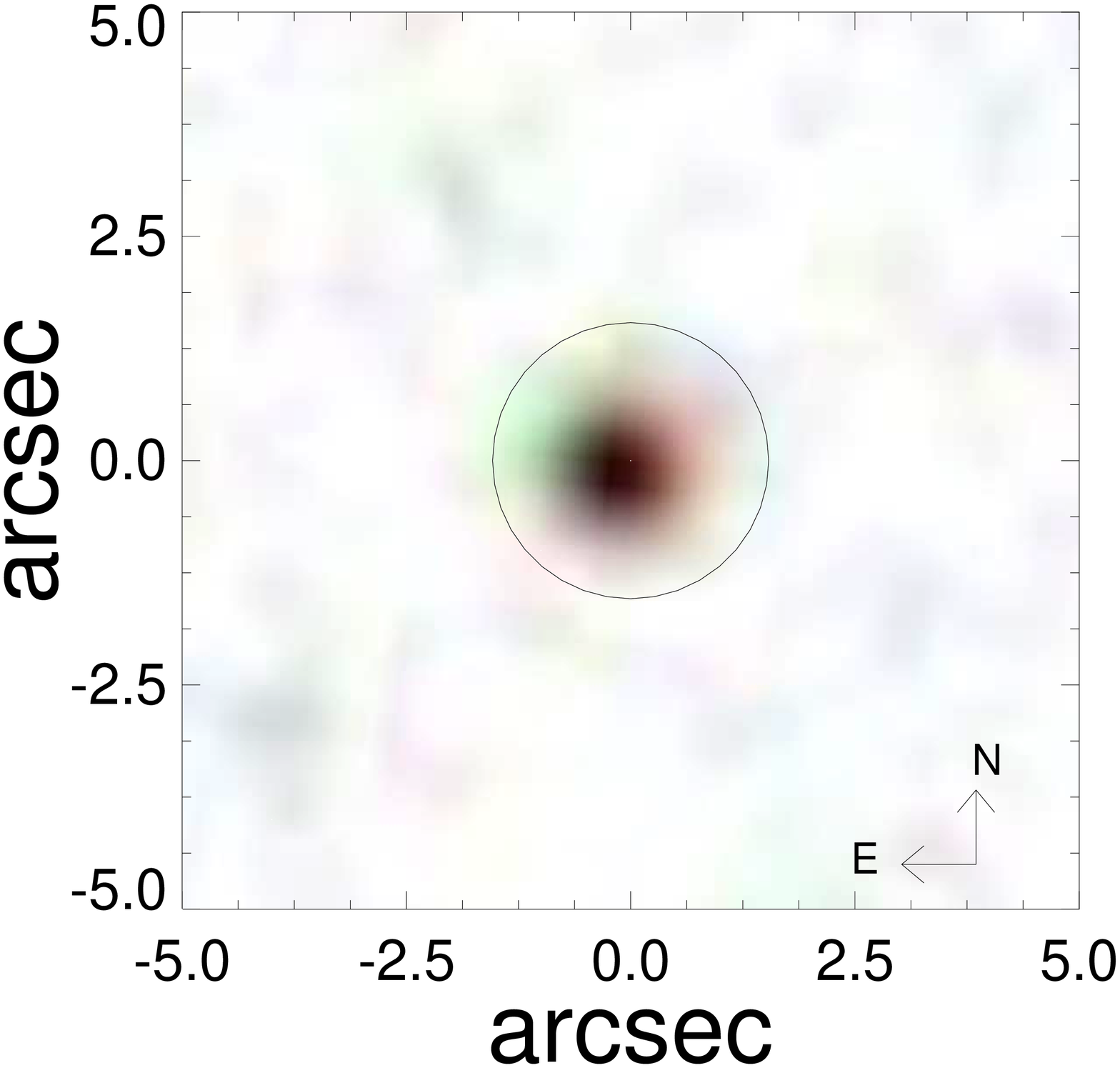,width=2.6cm,height=2.6cm,bbllx=-6bp,bblly=15bp,bburx=663bp,bbury=663bp,clip=true}\\
 \epsfig{figure=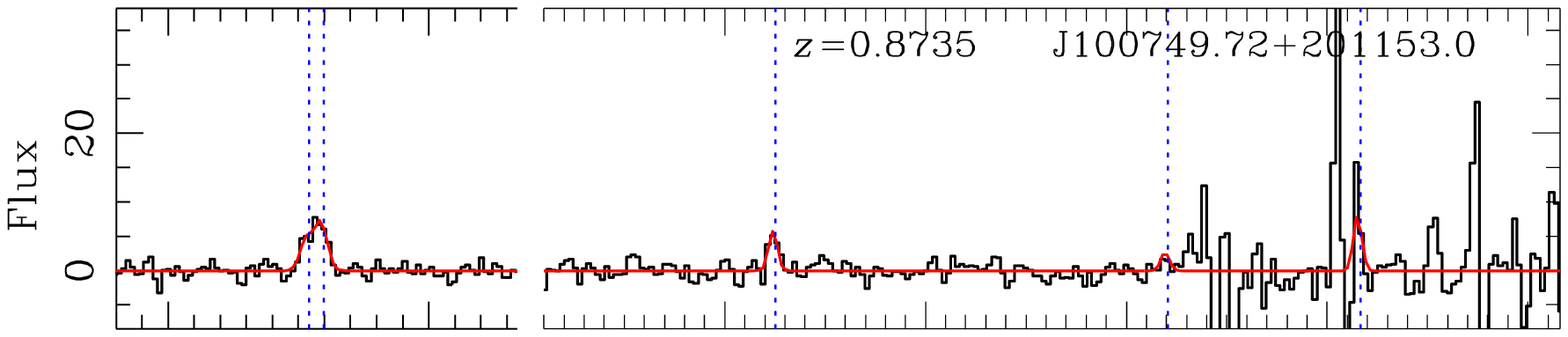,height=2.6cm,width=10.5cm,bbllx=32bp,bblly=190bp,bburx=550bp,bbury=325bp,clip=true} & \epsfig{figure=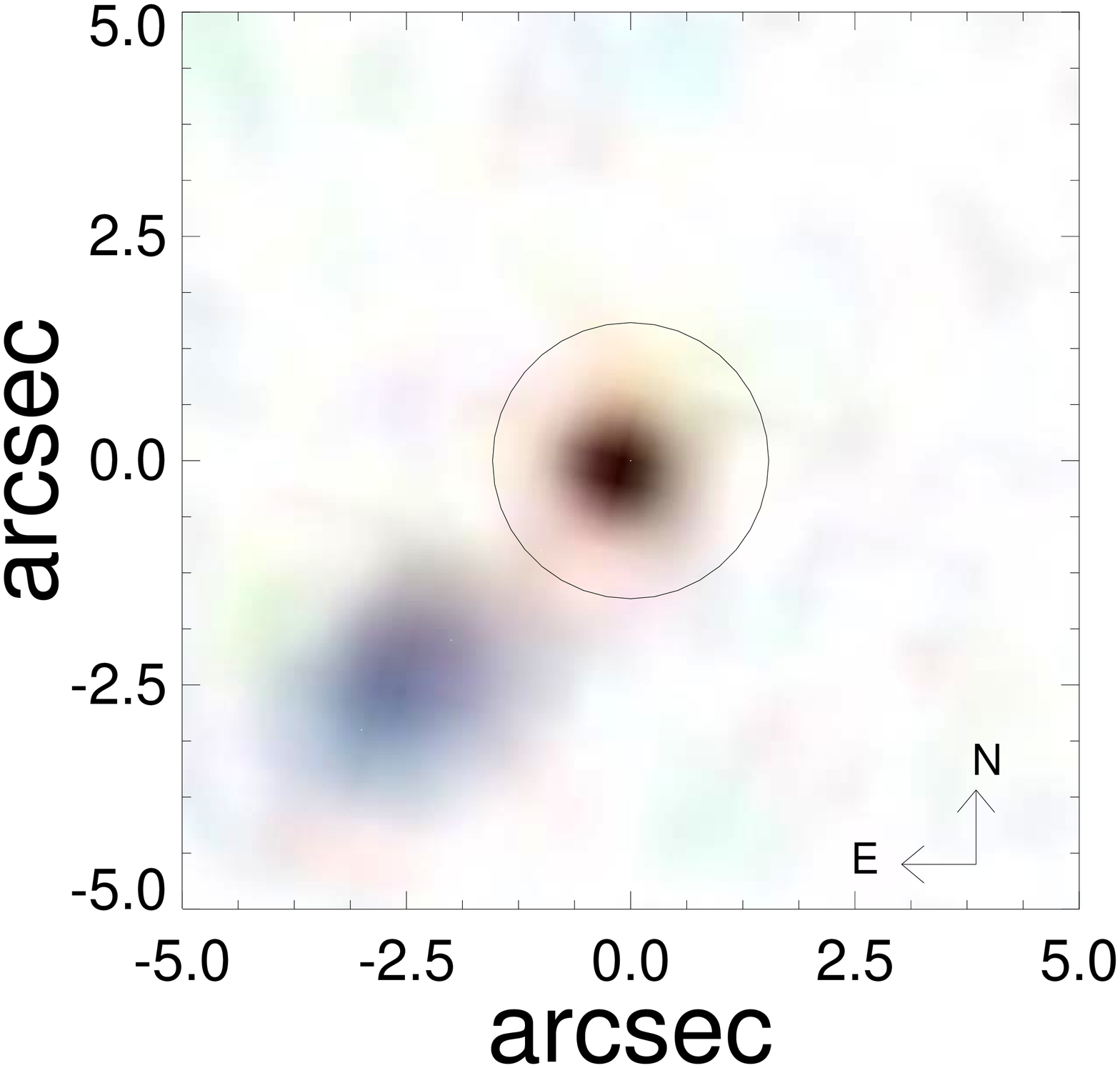,width=2.6cm,height=2.6cm,bbllx=-6bp,bblly=15bp,bburx=663bp,bbury=663bp,clip=true}\\

\epsfig{figure=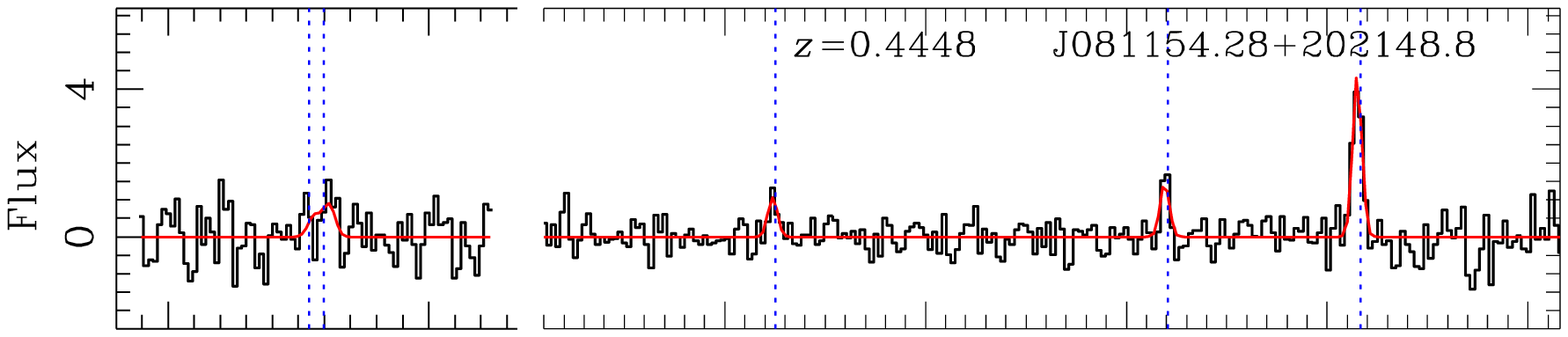,height=2.6cm,width=10.5cm,bbllx=32bp,bblly=190bp,bburx=550bp,bbury=325bp,clip=true} & \epsfig{figure=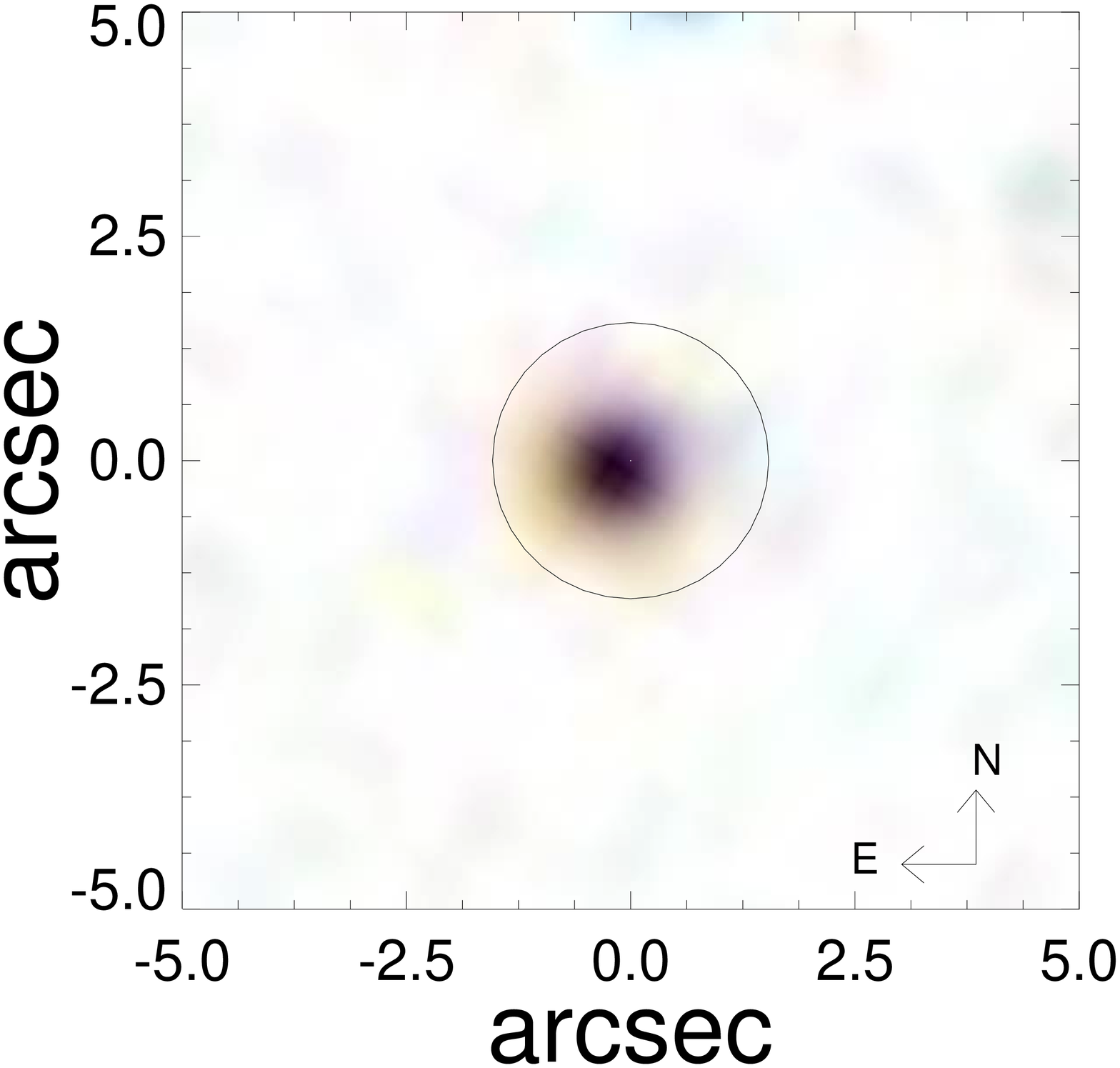,width=2.6cm,height=2.6cm,bbllx=-6bp,bblly=15bp,bburx=663bp,bbury=663bp,clip=true}\\
     \epsfig{figure=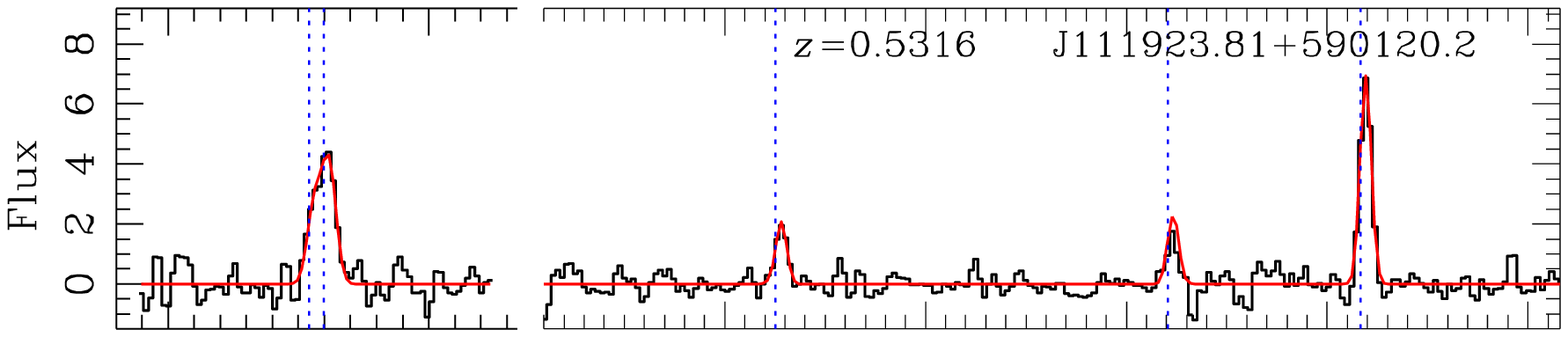,height=2.6cm,width=10.5cm,bbllx=32bp,bblly=190bp,bburx=550bp,bbury=325bp,clip=true} & \epsfig{figure=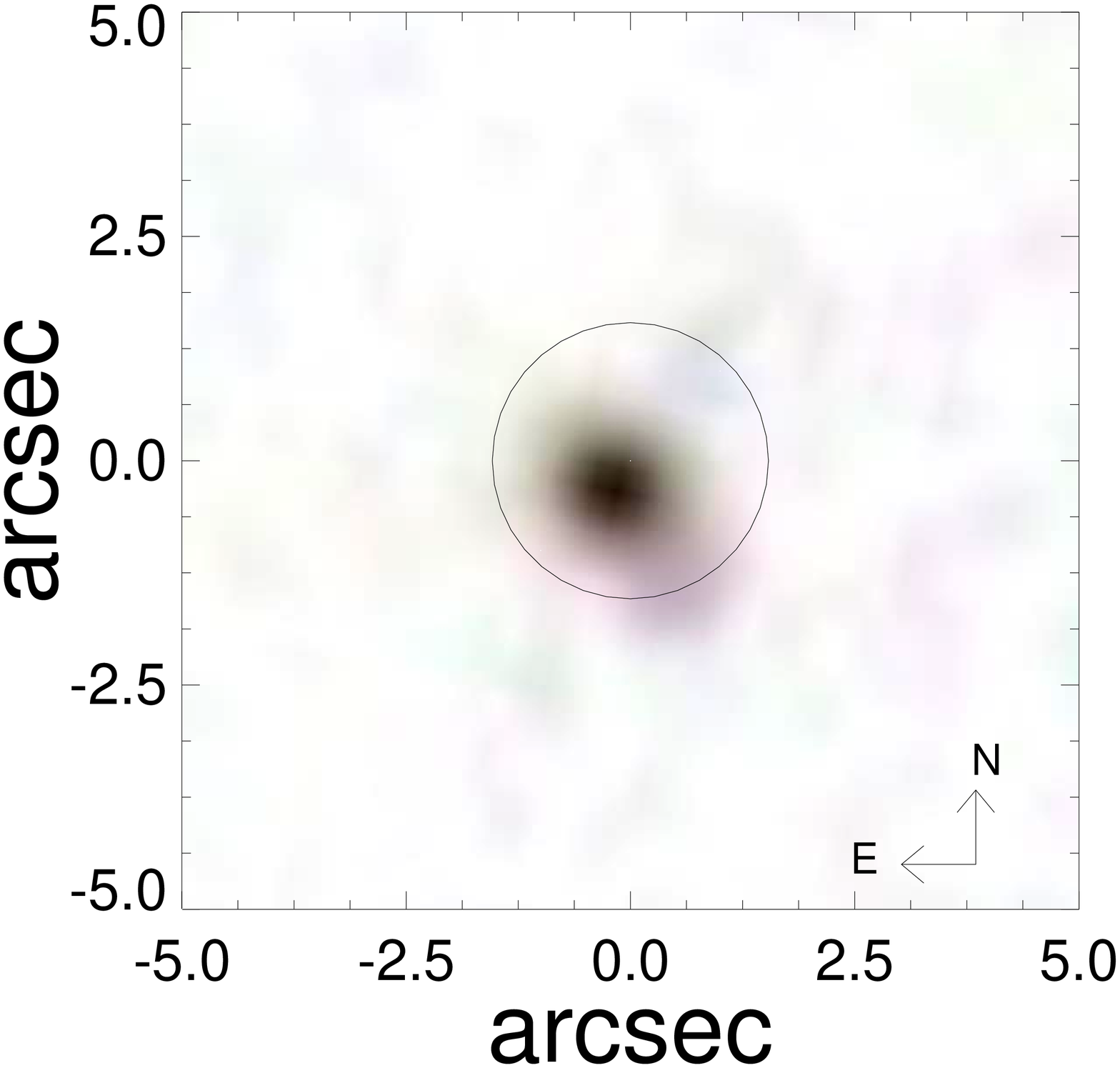,width=2.6cm,height=2.6cm,bbllx=-6bp,bblly=15bp,bburx=663bp,bbury=663bp,clip=true}\\

     \epsfig{figure=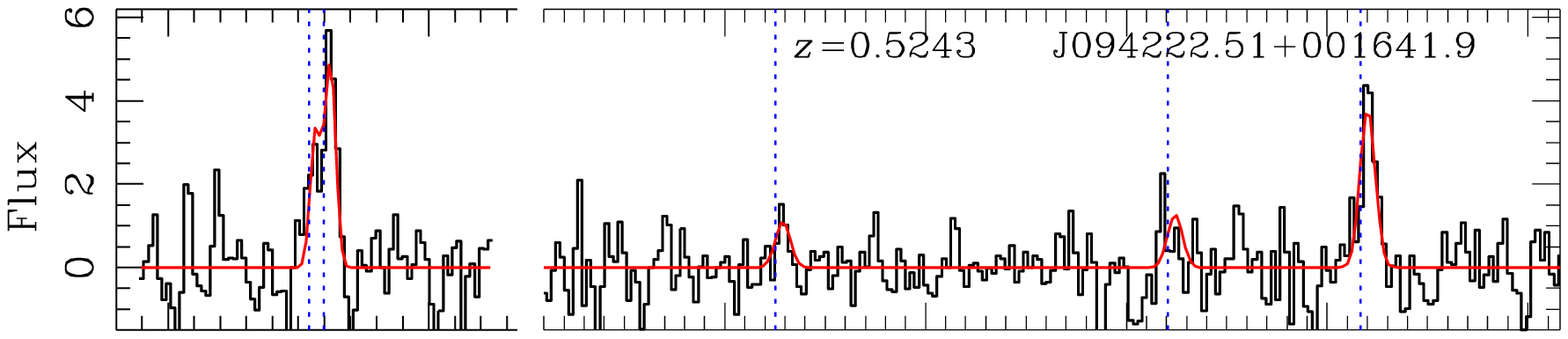,height=2.6cm,width=10.5cm,bbllx=32bp,bblly=190bp,bburx=550bp,bbury=325bp,clip=true} & \epsfig{figure=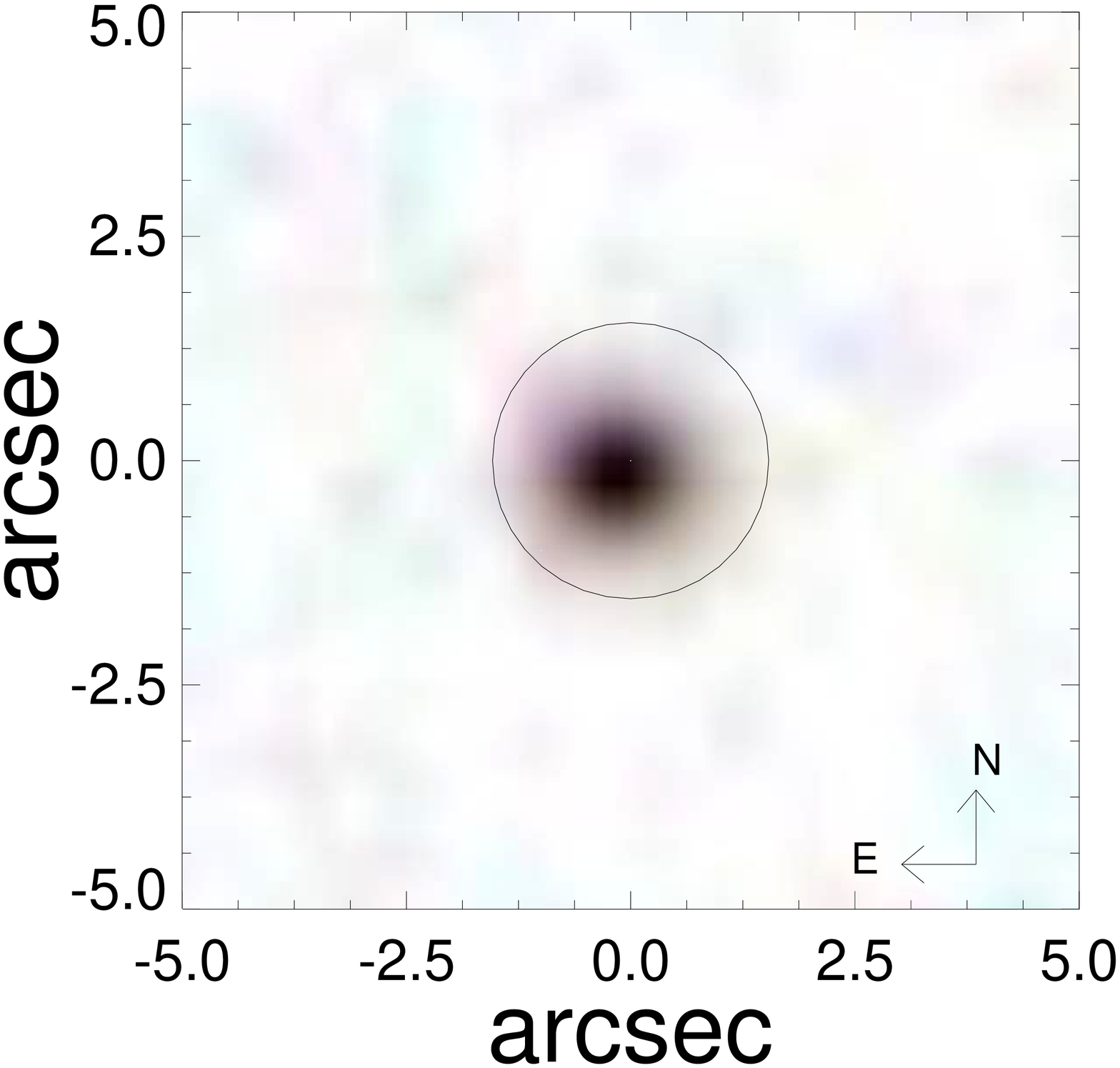,width=2.6cm,height=2.6cm,bbllx=-6bp,bblly=15bp,bburx=663bp,bbury=663bp,clip=true}\\
        \epsfig{figure=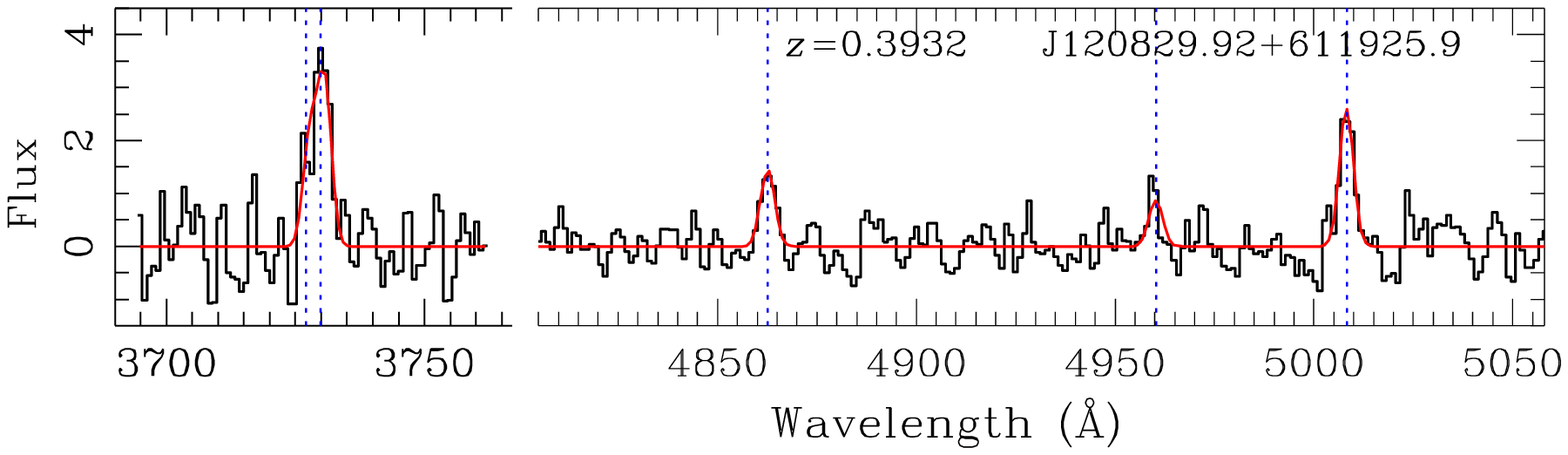,height=2.9cm,width=10.5cm,bbllx=32bp,bblly=170bp,bburx=550bp,bbury=325bp,clip=true} & \epsfig{figure=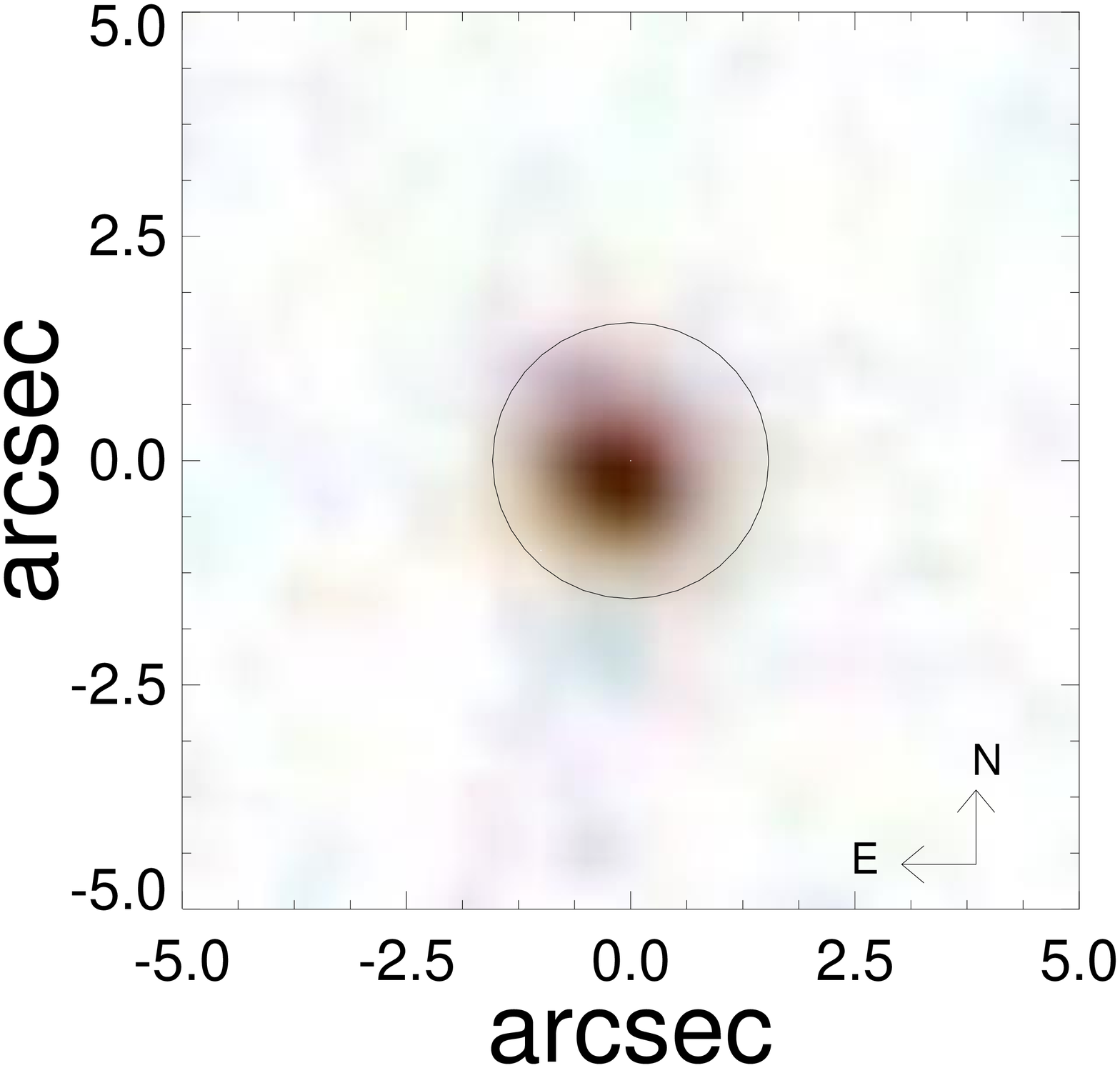,width=2.6cm,height=2.6cm,bbllx=-6bp,bblly=15bp,bburx=663bp,bbury=663bp,clip=true}\\
      \end{tabular}
      \caption{Examples of nebular emission lines detected in the
        continuum subtracted spectra from the intervening \mgii
        systems, the flux scale is in units of $\rm
        10^{-17}\ erg\ s^{-1}\ cm^{-2}\ \AA^{-1}$. In each panel QSO
        name, absorption redshift are indicated and vertical dashed
        lines indicates different nebular emission lines marked at the
        top panel. In the \emph{right hand panels} we show the image
        stamp of size $\sim 10$ arcsec around the quasars. The circle
        represents the position of the 3-arcsec-diameter SDSS-DR7
        fibre.}
      \label{fig:example}
      \end{center}
     \end{figure*}

However, identifications solely based on single emission line (i.e.,
blended \oiiab\ lines) alone may lead to a high probability of false
positives if the expected wavelength range is affected by (1) poor sky
subtraction in the locations of strong sky lines and/or (2) a bad
continuum fit around the broad and narrow emission lines of the
background quasar. Therefore, to avoid such cases we have masked the
regions around strong sky lines as well as narrow and broad emission
lines from QSOs. In addition, any feature with a full width at half
maximum (FWHM) less than the SDSS spectral resolution ($\sim
150$~\kms) are manually checked and then removed if found to be
spurious. We detect \oii\ emission from 185 \mgii systems in the
redshift range $0.35 \le$ \zabs $\le 1.1$ at $\ge~4\sigma$ level.
Among them, 62 are detected in SDSS-DR7 (with a fibre diameter of 3
arcsec) and remaining 123 are detected in SDSS-DR12 (with a fibre
diameter of 2 arcsec) spectra. From these \oii\ detected systems, we
also searched for the other emission features such as \hbetalam and
\oiiiab. Among 62 systems detected in SDSS-DR7, we found 29 systems
showing \oiiib and 13 systems showing \hbeta detected at $\ge
3\sigma$. In addition, among 123 systems detected in SDSS-DR12, we
found 76 systems showing \oiiib and 46 systems showing \hbeta detected
at $\ge 3\sigma$. The \zabs\ distributions of \mgii systems with the
nebular \oii\ emission detected in SDSS-DR7 and SDSS-DR12 are shown in
Fig.~\ref{fig:zabshist} .\par

In the general population of galaxies a small fraction (i.e $\sim$ 8
per cent) show the \oiii\ emission being stronger than \oii. The
nebular line selection based purely on \oii\ may miss such systems. In
order to account for such systems we searched for \oiii\ emission from
\mgii absorbers that do not show strong \oii\ emission (i.e at more
than 4$\sigma$ level). We found 13 \mgii systems where
\oiiiab\ emission is clearly detected at $\ge 3\sigma$ level while
\oiiab\ emission is not detected at the significant level demanded
above (i.e., $\ge 4\sigma$). Thus in total we have detected nebular
emission lines from 198 \mgii absorption systems.\par

In Fig.~\ref{fig:zabscomp} (\emph{top panel}), we compare the redshift
distribution of the \mgii systems with clear detection of both
\oii\ and \oiii\ emission versus the systems with only
\oii\ detection. This figure (top panels) also shows the $z$
distribution of 13 \oiii\ detected systems with weak \oii. It is clear
from the figure that systems with only \oii\ detection are
preferentially at high redshift. A two-sided Kolmogorov-Smirnov test
($KS-$test) shows that the two redshift distributions (i.e., with both
\oii, \oiii\ detections and only \oii\ detections) are different with
a null probabilities of being drawn from same parent distribution
$p_{null} =$ 0.03 and 0.04 for the systems detected in SDSS-DR7 and
SDSS-DR12, respectively. We find that the \oii\ luminosity (total
luminosity of two \oii\ lines) for these two subsets are similar (see
\emph{bottom panels} of Fig.~\ref{fig:zabscomp}) with $KS-test$ null
probability of $p_{null} =$ 0.42, 0.62 for the SDSS-DR7 and SDSS-DR12,
respectively. The redshift distribution of \oiii\ detected systems are
also different, preferentially at low-$z$. These differences could
either mean (i) some observational bias against detecting
\oiii\ emission from high-$z$ galaxies, for example the poor sky
residuals or (ii) a redshift evolution in the \o3o2 ratio. We come
back to this is Section~\ref{sub:o3o2}. The details of our sample are
summarized in Table~\ref{tab:sample}. The final list of systems with
nebular emission is available in the online table. In all cases where
we do not have nebular line (i.e., one of \oiii, \hbeta or \oii)
detections we have estimate the $3\sigma$ upper limit by assuming the
FWHM of the line that is detected.

 \begin{table*}
 \centering
 \begin{minipage}{160mm}
 {\scriptsize
 \caption{Sample of \mgii absorbers with  nebular emission line$^{\textcolor{blue}{a}}$.}
 \label{tab:sample}
 \begin{tabular}{@{} c c r c r c c r r r r @{}}
 \hline  \hline 
  \multicolumn{1}{c}{No}   &\multicolumn{1}{c}{QSO}  &       \multicolumn{1}{c}{plate} & \multicolumn{1}{c}{mjd} &\multicolumn{1}{c}{fibre}& \multicolumn{1}{c}{zqso} & \multicolumn{1}{c}{zabs} & \multicolumn{1}{c}{\loii} & \multicolumn{1}{c}{\loiii}  & \multicolumn{1}{c}{L$_{\rm H \beta}$} & \multicolumn{1}{c}{\ew}  \\ 
  \multicolumn{1}{c}{}   &\multicolumn{1}{c}{   }  &       \multicolumn{1}{c}{     } & \multicolumn{1}{c}{   } &\multicolumn{1}{c}{     }& \multicolumn{1}{c}{    } & \multicolumn{1}{c}{    } & \multicolumn{3}{c}{($\rm \times 10^{40} erg\ s^{-1}$)} & \multicolumn{1}{c}{(\AA)}\\

\hline 
     1 &  BOSS-J$\rm000910.01+110715.6$&   6113&  56219&  602&    2.6470&    0.6804&     21.69$\pm$ 2.21&  14.78$\pm$ 2.78            &     10.96$\pm$     2.11&  2.92$\pm$0.17      \\
     2 &  BOSS-J$\rm004427.51+152439.2$&   6198&  56211&  086&    2.6090&    0.7142&     22.03$\pm$ 2.76&  16.76$\pm$ 2.66            &     $<6.43$            &  2.77$\pm$0.61 \\
     3 &  SDSS-J$\rm020317.22-010122.8$&   0701&  52179&  087&    1.7210&    0.7253&     26.10$\pm$ 7.53&   $<31.89^{\textcolor{blue}{b}}$     &  $<32.67$   &2.10 $\pm$ 0.51\\  
     4 &  SDSS-J$\rm023618.98-000529.1$&   0408&  51821&  237&    0.9806&    0.6448&     17.16$\pm$ 4.86&   $<13.70$                  &  $<11.58$   &1.26 $\pm$ 0.40\\  
     5 &  SDSS-J$\rm030730.61-074555.6$&   0459&  51924&  309&    0.7543&    0.6986&     28.81$\pm$ 4.00&     45.09 $\pm$      8.49   &  $<13.40$   &2.34 $\pm$ 0.27\\  
     6 &  SDSS-J$\rm074850.59+165625.5$&   1920&  53314&  217&    0.9321&    0.6651&     21.86$\pm$ 4.20&   $<15.02$                  &$12.82\pm4.56$&2.07 $\pm$ 0.26\\  
     7 &  SDSS-J$\rm081121.39+451719.4$&   0439&  51877&  554&    0.8907&    0.6702&     20.42$\pm$ 3.92&   $<21.16$                  &  $<17.61$   &0.92 $\pm$ 0.09\\  
     8 &  SDSS-J$\rm084507.04+311003.0$&   1270&  52991&  266&    1.0903&    0.6370&     35.03$\pm$ 8.22&   $<12.12$                  &  $<18.77$   &0.95 $\pm$ 0.12\\  
     9 &  SDSS-J$\rm085051.97+083026.6$&   1299&  52972&  333&    0.6635&    0.3767&      5.56$\pm$ 1.74&      2.82 $\pm$      0.88   &  $ <2.16$   &1.75 $\pm$ 0.15\\  
    10 &  SDSS-J$\rm085846.71+531643.2$&   0449&  51900&  145&    1.1859&    0.7886&     42.37$\pm$ 9.69&   $<93.84$                  &  $<25.97$   &1.77 $\pm$ 0.08\\  
\hline
 \end{tabular} 
 }  \\          
    $^{\textcolor{blue}{a}}$ The full table is available online. \\
    $^{\textcolor{blue}{b}}$ $3\sigma$ upper limit.

 \end{minipage}
 \end{table*}

Note that, our compilation of 198 \mgii systems, seen in emission, is
the largest sample of \mgii galaxies known till date. Based on
previous attempts to study the galaxies associated with \mgii
absorbers only $\sim$ 182 spectroscopically identified intermediate
redshift ($0.07 \le z \le 1.1$) galaxies at impact parameters of $5.4
\le \rho \le 193$ kpc are known (see the compilation of \citet[][and
  reference therein]{Nielsen2013ApJ...776..114N}. In particular, the
\mgii systems with nebular emission in our compilation probe very low
impact parameters (i.e. $\rho \le 12$ kpc). In the literature only 6
galaxies associated with \mgii absorbers are known at such impact
parameters. In addition, nebular emission from 17 \mgii systems are
identified by \citet{Noterdaeme2010MNRAS.403..906N} in the SDSS fibre
spectra. Eight of these systems are part of our sample.

\subsection{Emission line parameters}
We determine the emission line parameters by fitting the \oiiab,
\hbetalam and \oiiiab, emission lines with Gaussian profiles. We fit
all detected lines simultaneously by using a single emission redshift
within a velocity offset of $\pm 150$~\kms\ with respect to the
absorption redshift as typically seen in GOTOQs
\citep[see,][]{Noterdaeme2010MNRAS.403..906N}. In the low resolution
SDSS spectrum the two \oiiab\ lines are unresolved. We model the
observed \oii\ emission with a double Gaussian profile with a tied
linewidth but freely varying the line ratio in a range of
3.4-1.5{\footnote{The {\sc [O II]$\lambda3729/\lambda3727$} intensity
    ratio in the range 3.4-1.5 is predicted in photoionization models
    for the electron density in the range ${\it n_e = \rm 10^1-10^5
      cm^{-3}}$ for the kinetic temperature $T = 10,000^{\circ} \rm K$
    \citep{Osterbrock2006agna.book.....O}.}}. This allows for probing
the typical range in the electron density of the gas under
photoionization equilibrium. The \hbeta and \oiiiab\ emission lines
are modelled with a single Gaussian each. The integrated line
intensities of \oiiab, \oiiiab\ and \hbeta are then measured from the
fitted Gaussian parameters.

 \begin{figure}
 \epsfig{figure=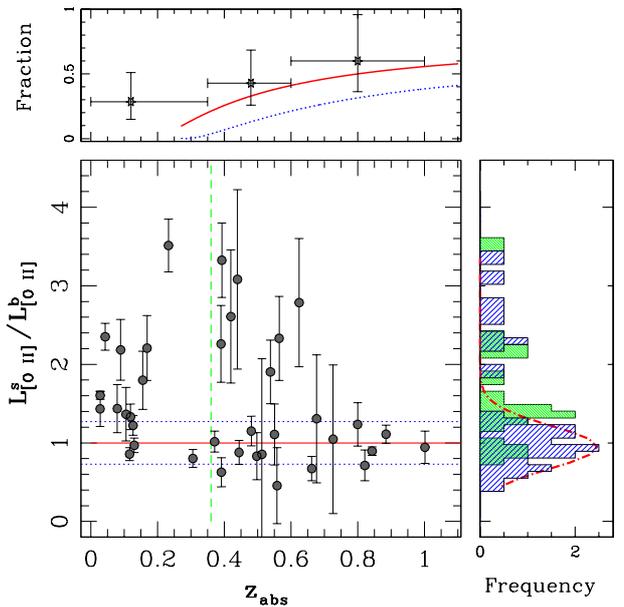,height=8.2cm,width=8.2cm}
  \caption{ The \oii\ luminosity ratio measured between SDSS-DR7 and
    SDSS-DR12 spectra (i.e., L$_{[\rm O II]}^s$/L$_{[\rm O II]}^b$) as
    a function of absorber redshift. The right panel shows the
    histogram of L$_{[\rm O II]}^s$/L$_{[\rm O II]}^b$ for the systems
    detected at high (i.e., $z > 0.36$) redshift (\emph{solid
      histogram, shaded with slanted lines at 45$^{\circ}$}) and
    fitted with Gaussian function and at low (i.e. $z < 0.36$)
    redshift (\emph{dashed histogram, shaded with slanted lines at
      -45$^{\circ}$}). The top panel shows the fraction of systems with
    \loii\ consistent within $1\sigma$ range over three redshift bins.
    The \emph{solid} (respectively, \emph{dotted}) curve represents
    the expected $z$ dependence of absorber galaxy being inside the
    SDSS-DR7 (respectively, SDSS-DR12) fibre (see
    Section~\ref{lab:fibre_effect} for details).}
\label{fig:fibre_ratio}
 \end{figure}

In Fig.~\ref{fig:example}, we present few examples of the nebular
emission line spectra of \mgii systems in our sample along with the
results of our best fit Gaussian profiles. The lower three panels show
examples of clear detections of \oii, \hbeta and \oiii\ emission
lines. In upper two panels we show examples where \oii\ is clearly
detected whereas the \oiii\ region is affected by the night sky
emission. In addition, third panel from the top shows an example of
GOTOQ system where \oiii\ emission is clearly seen without a clear
detection of the \oii\ line.

\begin{figure}
 \epsfig{figure=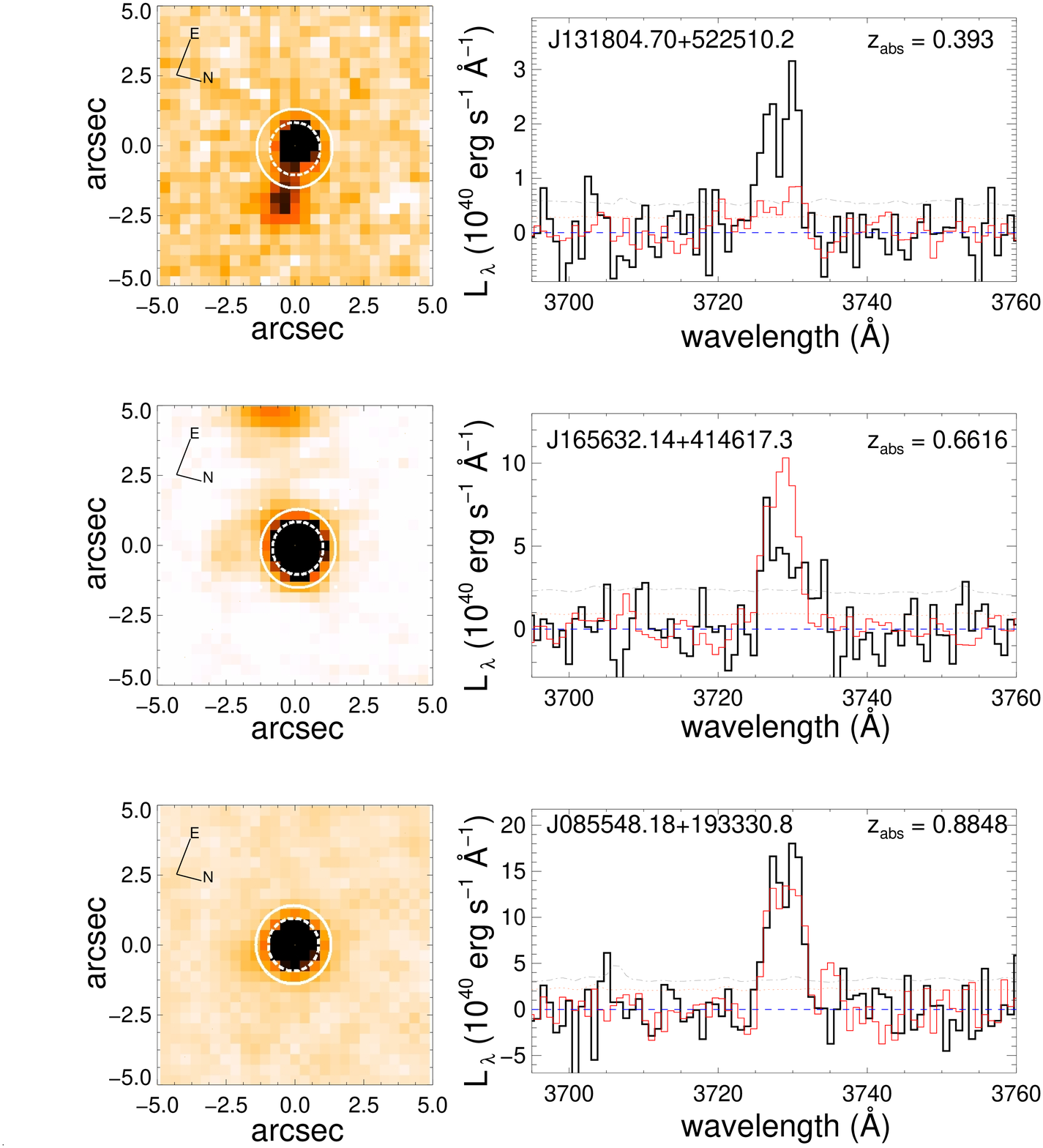,height=8.2cm,width=8.2cm,bbllx=43bp,bblly=0bp,bburx=922bp,bbury=1009bp,clip=true}
  \caption{ \emph{Right panels:} Examples of variation in
    \oii\ nebular emission line luminosity with change in fibre size
    from 3 to 2 arcsec in SDSS-DR7 (\emph{thick solid line}) and DR12
    (\emph{thin solid line}) spectrum, respectively. The error spectra
    of SDSS-DR7 and SDSS-DR12 are shown as \emph{dot-dashed} and
    \emph{dotted} lines. \emph{Left panels:} Image stamp of size
    $\sim$ 10 arcsec around the quasar. The \emph{dotted} and
    \emph{solid} circles represent the size corresponding to 2 and 3
    arcsec-diameter of SDSS-DR12 and SDSSS-DR7 fibre, respectively.
    Note that, the fibre centering may be slightly different than what
    is shown here. }
\label{fig:fibre_exam}
 \end{figure}

\section{fibre size effect} 
\label{lab:fibre_effect}

Our sample of \mgii systems that show nebular emission lines spans a
substantial redshift interval, ranging from 0.4 to 1.1. In this range,
the fibre with an aperture of 3 arcsec, used in SDSS-DR7 spectra,
corresponds to a projected radius (i.e., an upper limit on the impact
parameter) of $\sim 8.1-12.3\ \rm kpc$. Similarly, the 2 arcsec fibre
used in SDSS-DR12 corresponds to a radius of $\sim 5.4-8.2\ \rm kpc$.
Before performing different statistical analysis it is important for
us to understand various possible biases introduced by the finite size
of fibres used. For this we used the repeated spectroscopic
observations of quasars in SDSS-DR7 and SDSS-DR12. Including the
GOTOQs listed in \citet{Straka2015MNRAS.447.3856S} we have found 39
cases with repeat observations using fibre of two different sizes. \par

  In Fig.~\ref{fig:fibre_ratio}, we compare the ratio of
  \oii\ luminosities measured between SDSS-DR7 (i.e., L$_{[\rm O
      II]}^s$) and SDSS-DR12 (i.e., L$_{[\rm O II]}^b$) fibre spectra
  as a function of the absorber redshift. For this comparison, we
  consider only systems where the nebular emission line is clearly
  detected at least in one epoch spectrum. As expected the measured
  luminosities in the SDSS-DR7 spectra are higher than those of
  SDSS-DR12 spectra in several cases. Contrary to our expectation 18\%
  of \mgii systems at $z > 0.4$ show slightly larger luminosity in the
  SDSS-DR12 spectrum albeit within $\sim 2\sigma$ level. This is
  possible when there are issues related to centering of the quasars
  in the fibre during spectroscopic observations or effect of varied
  seeing between two epochs. In addition, there could also be issues
  related to flux scales, where SDSS-DR12 spectra tend to show excess
  flux in the blue. We consider the scatter in \oii\ luminosity ratio
  as the level at which we can not quantify the flux difference over
  the redshift range of our interest (i.e., $0.35 \le$ \zabs $\le
  1.1$), shown as dotted lines in Fig~\ref{fig:fibre_ratio}. It is
  interesting to note that the fluxes in the SDSS-DR12 and SDSS-DR7
  agree within the 1$\sigma$ uncertainty in most of the cases. It is
  clear from Fig.~\ref{fig:fibre_ratio} that the fibre size effect is
  severe at low-$z$. Considering the $1\sigma$ uncertainty, the
  fraction of systems with similar fluxes in SDSS-DR7 and DR12 is
  found to be 8\%, 42\% and 50\% for the three redshift range of $z <
  0.35$, $0.35 \le z \le 0.6$ and $z \ge 0.6$ (see top panel of
  Fig.~\ref{fig:fibre_ratio}). In Fig.~\ref{fig:fibre_exam}, we show
  three examples demonstrating the fibre size effect where the
  \oii\ luminosity seen in SDSS-DR7 (\emph{thick solid line}) spectrum
  is found to be larger (\emph{top panel}), smaller (\emph{middle
    panel}) and similar (\emph{bottom panel}) to those measured in the
  SDSS-DR12 (\emph{thin solid line}) spectrum. \par

\begin{figure}
 \epsfig{figure=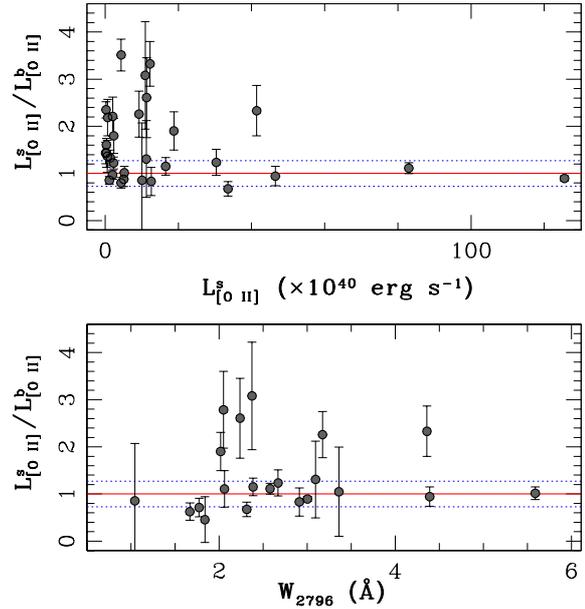,height=8.2cm,width=8.2cm}
  \caption{\emph{Top panel:} The \oii\ luminosity ratio measured in
    SDSS-DR7 and DR12 spectra as a function of \oii\ luminosity
    detected in SDSS-DR7 spectra. \emph{Bottom panel:} The
    \oii\ luminosity ratio as a function of \ew.}
\label{fig:fibre_vs_loii}
 \end{figure}

 Next, in Fig.~\ref{fig:fibre_vs_loii} we plot the \oii\ luminosity
 ratio (i.e., L$_{[\rm O II]}^s$/L$_{[\rm O II]}^b$) against the
 \loii\ measured in SDSS-DR7 spectra, i.e., L$_{[\rm O II]}^s$, and
 \ew . It is clear from the top panel of Fig.~\ref{fig:fibre_vs_loii}
 that a large scatter in L$_{[\rm O II]}^s$/L$_{[\rm O II]}^b$ is
 preferentially present for systems with low \loii. It is quite
 possible that the line emitting region is partially covered by the
 fibre which leads to the lower \loii. This figure also demonstrates
 that the observed \oii\ luminosity in the SDSS-DR12 spectra could be
 underestimated by up to a factor of 3 due to fibre size effects.

 Recently, \citet{Paulino-Afonso2017MNRAS.465.2717P} have found the
 effective radius ($r_{\rm e}$) of H$\alpha$ selected galaxies
 (relevant for the present study) shows a mild decrease over
 increasing $z$. They measured a median size of $r_{\rm e}$ $\sim 4$
 kpc at $z \sim 0.4$. If we assume this size for the nebular line
 emitting regions studied here then at $z \sim 0.4$ all
 \oii\ luminosity from the galaxies within $r_{\rm e}$ will be
 observed only when $\rho < 4.1$~kpc and $< 1.4$ kpc in the case of
 SDSS-DR7 and SDSS-DR12, respectively. Therefore, if we assume $\rho$
 of our detections to be uniformly distributed within the fibre then
 only in $\le 25\%$ and $\le 7\%$ cases we expect fibre loss to be
 negligible in the case of SDSS-DR7 and SDSS-DR12 spectra respectively
 at $z \sim 0.4$. However, at $z \sim 0.8$ the median $r_{\rm e} \sim
 3.3$~kpc. In this case we will detect full \oii\ nebular emission in
 $\sim 50\%$ cases for SDSS-DR7 and $\sim 32\%$ cases for SDSS-DR12.
 The probability of \mgii galaxy in our sample not suffering the fibre
 loss, measured as (($r_{\rm fibre} - r_{\rm e})/r_{\rm fibre})^2$, as
 a function of $z$ for SDSS-DR7 (\emph{solid curve}) and SDSS-DR12
 (\emph{dotted curve}), are shown in the top panel of
 Fig.~\ref{fig:fibre_ratio}. Here, $r_{\rm fibre}$ is the projected
 radius of the fibre at the redshift of the \mgii systems. We consider
 this as an upper limit as a detection can occur even when the impact
 parameter is slightly larger than the fibre (i.e. when $r_{\rm fibre}
 \ge \rho - r_{\rm e}$).  \emph{It is clear from the above
     discussions that our measurements of nebular line luminosities
     are affected by fibre size effects and this effect has  a clear redshift
     dependence.}

\section{RESULTS}

In what follows we study the emission line properties of \mgii
absorbers and compare these with the properties derived using
absorption lines.

\subsection{Absorption line properties:}

In order to see the fibre size effect on the absorption line
properties, we first compare \ew, \mgii doublet ratio ($DR=W_{\rm Mg~II \lambda2796}/W_{\rm Mg~II \lambda2803}$) and $W_{\rm Fe~II
  \lambda2600}/W_{\rm Mg~II \lambda2796}$ (defined as $R$) of the
\mgii absorbers with \oii\ detections in SDSS-DR7 and DR12 data set.
Naively we would have expected the average \ew\ to be slightly lower
in the case of SDSS-DR7 as it also could have sightlines with slightly
larger impact parameters due to larger fibre size. A two-sided
Kolmogorov-Smirnov test finds no difference between the two subsets
based on \ew, $DR$ and $R$ parameters with a null probability of being
drawn from same parent distribution of $P_{KS} = 0.4, 0.2\ \rm
and\ 0.4$, respectively. The lack of any significant difference can be
attributed to the flattening of \ew\ vs $\rho$ and possible scatter in
this relationship. We come back to this in Section~\ref{sub:impact}.

Next, we compare the distribution of $DR$ and $R$ of the \mgii systems
with and without detection of the nebular emission lines. At first, we
found that all but 8 \mgii systems detected in emission (both in DR7
and DR12) have \ew\ $\ge 1$~\AA, with mean \ew\ of $\sim$ 2.3~\AA\ and
2.4~\AA\ for the SDSS-DR7 and SDSS-DR12, respectively. In
Fig.~\ref{fig:dndzdw_dr12} (\emph{lower panel}), we show the
distribution of \ew\ from the \mgii systems with detected nebular
emission line with the overall \mgii absorbers in our primary sample.
It is clear from the figure that \mgii systems with nebular emission
are predominantly distributed towards higher \ew.

 Using deep imaging of 7 \mgii systems with \oii\ nebular emission at
 $z \sim 0.1$ with $\rho < 6$ kpc \citet{Kacprzak2013ApJ...777L..11K}
 have argued that these systems are consistent with \ew\ distribution
 of the Milky Way (MW) interstellar medium (ISM) and ISM+Halo.
 Following their approach, in the \emph{top panel} of
 Fig.~\ref{fig:dndzdw_dr12}, we compare the \ew\ distribution of
 systems with nebular emission in our sample with those measured from
 different components of MW. For this, we use 71 \mgii absorption
 lines produced by MW (ISM+halo gas) and 21 systems as halo gas
 identified along the sightlines of 83 quasars, observed in Hubble
 Space Telescope Quasar Absorption Line Key Project
 \citep{Bahcall1993ApJS...87....1B}, by
 \citet{Savage2000ApJS..129..563S}. The \ew\ distribution for the
 \mgii systems with \oii\ emission, MW(ISM+halo) and MW(halo gas) is
 shown in the \emph{upper panel} of Fig.~\ref{fig:dndzdw_dr12}. In
 view of the fact that MW sightlines pass halfway through the
 disk/halo we have applied a correction factor of two in the \ew\ for
 MW systems \citep{Kacprzak2013ApJ...777L..11K}. This figure clearly
 indicates that for a large fraction of \mgii systems with
 \oii\ emission the \mgii absorption tend to be strong and their
 distribution is closer to what has been seen for MW (ISM+halo). A
 simple two parameter $KS-$test confirms this.

\begin{figure}
 \epsfig{figure=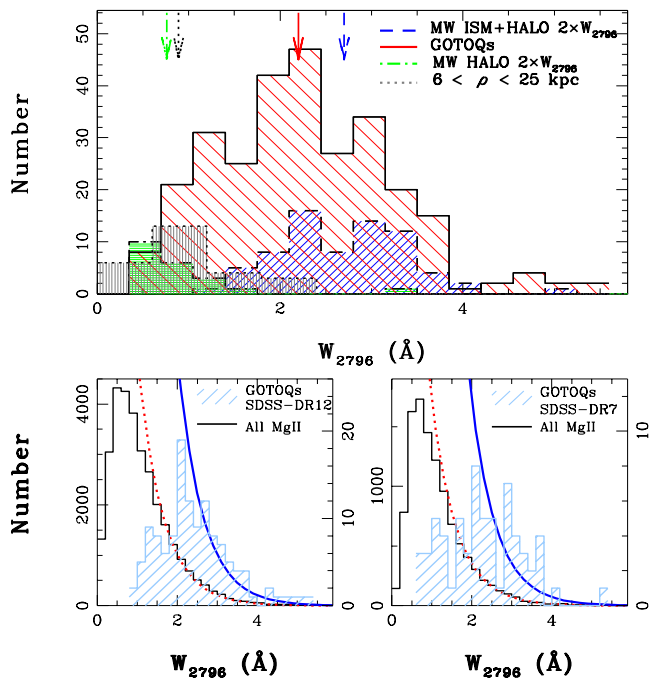,height=9.0cm,width=8.5cm,bbllx=19bp,bblly=166bp,bburx=403bp,bbury=585bp,clip=true}
\caption{\emph{Upper panel:} A comparison of the \mgii equivalent
  width (\ew) distribution of GOTOQs (\emph{solid histogram}) with the
  \mgii systems having $6 < \rho < 25$ kpc (\emph{dotted histogram,
    shaded with vertical lines}) and those through the Milky Way [MW]
  ISM+halo (\emph{dashed histogram, shaded with slanted lines at}
  45$^{\circ}$) and MW halo (\emph{dot-dashed histogram, shaded with
    horizontal lines}). For the MW systems a correction factor of two
  is applied as sight lines are intercepted halfway through the
  disk/halo. The median \ew\ for each sample is marked as an arrow on
  top. \emph{Lower left panel:} Distribution of \mgii rest equivalent
  widths for the overall SDSS-DR12 \mgii sample (with \zabs =
  $0.3-1.1$, unfilled histogram) and the \mgii absorbers that show
  nebular emission (hatched histogram). The two distributions are
  represented with different scales for presentation purpose only
  (SDSS in the left, galaxy sample in the right ordinates). The
  \emph{dotted} and \emph{solid} curves represent the parametrization
  by \citet{Zhu2013ApJ...770..130Z}, scaled to match the number of
  systems with \ew$\ge 2$~\AA. \emph{Lower right panel:} same as left
  for the SDSS-DR7 \mgii absorbers.}
 \label{fig:dndzdw_dr12}
 \end{figure}

\begin{figure}
 \epsfig{figure=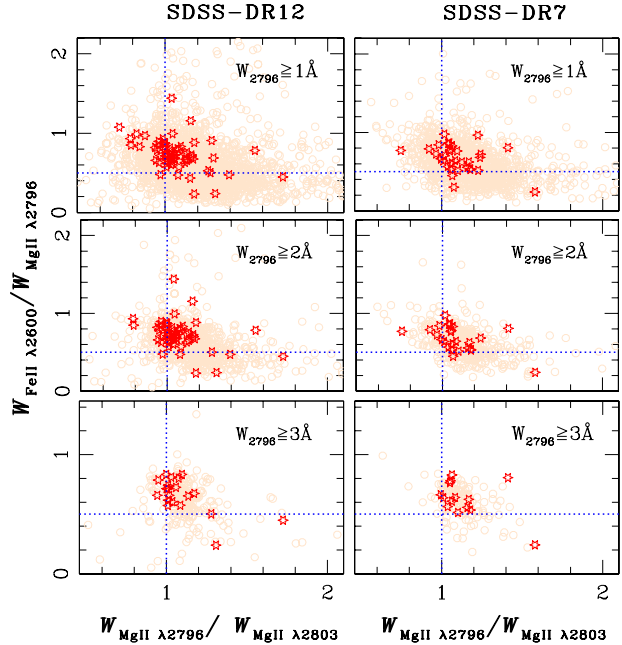,height=8.5cm,width=8.5cm}
\caption{The \mgii doublet ratio versus R parameter of \mgii absorbers
  with and without nebular emission line detection for \ew $\ge
  1$~\AA\ (\emph{top}), $\ge 2$~\AA\ (\emph{middle}) and $\ge 3$~\AA
  (\emph{bottom}) in the SDSS-DR12 (\emph{left panel}) and SDSS-DR7
  (\emph{right panel}) samples. \emph{Open circles} are for the full
  sample and \emph{red stars} are for \mgii systems with nebular
  emission.}
\label{fig:scatter}
 \end{figure}

\begin{figure}
 \epsfig{figure=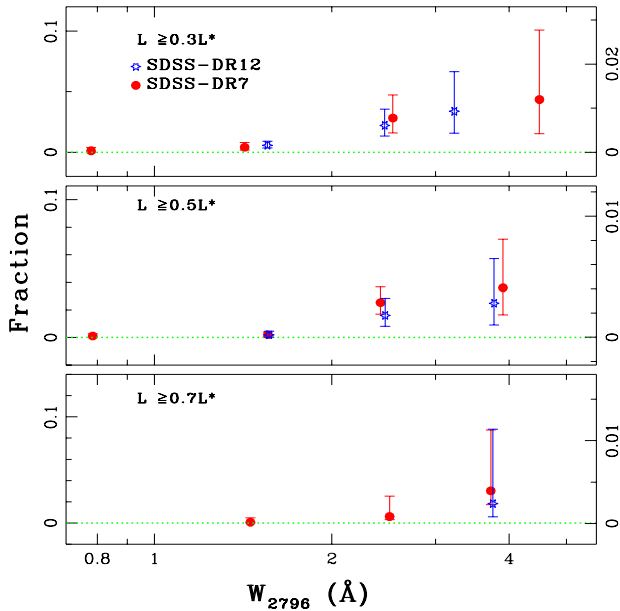,height=8.2cm,width=8.8cm}
\caption{Plot showing the fraction of \mgii systems detected in emission,
  with a luminosity threshold of $\rm \ge 0.3$ \lsoii (\emph{top
    panel}), $\rm \ge 0.5$ \lsoii (\emph{middle panel}) and $\rm \ge
  0.7$ \lsoii (\emph{bottom panel}) detected in SDSS-DR7
  (\emph{circle}) and SDSS-DR12 (\emph{stars}) fibre spectra. The
  fraction for \mgii absorbers in SDSS-DR12 fibre spectra is given in
  the right side ordinates. }
\label{fig:fraction}
 \end{figure}

\begin{figure}
 \epsfig{figure=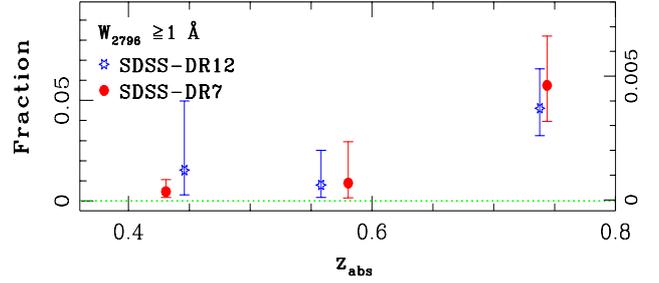,height=3.8cm,width=8.5cm,bbllx=27bp,bblly=496bp,bburx=576bp,bbury=712bp,clip=true}
\caption{Plot showing the fraction of strong \mgii systems (\ew $\ge
  1$~\AA) detected in emission, with a luminosity threshold of $\rm
  \ge 0.3$ \lsoii\ detected in SDSS-DR7 (\emph{circle}) and SDSS-DR12
  (\emph{stars}) fibre spectra, as a function of $z$. The fraction for
  \mgii absorbers in SDSS-DR12 fibre spectra is given in the right
  side ordinates. }
\label{fig:fraction_vs_z}
 \end{figure}

Using a photoionization modelling \citet{Srianand1996ApJ...462..643S}
has shown that the \mgii absorbers with $R \ge 0.5$ trace systems with
high $N$(H{\sc~i}), preferentially the Damped \lya absorbers (DLAs)
\citep[see
  also,][]{Rao2006ApJ...636..610R,Gupta2012A&A...544A..21G,Dutta2017MNRAS.465.4249D}.
Furthermore, \citet{Rao2006ApJ...636..610R} have shown that the
fraction of \mgii systems that are DLAs increases with the \ew. In
order to find if the \mgii absorbers that show nebular line emission
occupy any preferred location in the parameter space defined by the
equivalent width ratios of metal absorption, we plot the $DR$ versus
$R$ in Fig.~\ref{fig:scatter}, for systems with and without nebular
emission detection over various \ew\ bins (i.e., $\rm \ge 1 \AA, \ge 2
\AA$ and $\ge$ 3~\AA). It is clear from the figure that among the
strong \mgii absorbers the systems showing nebular emission are
preferentially having a small $DR$, close to unity, and the $R$
parameter greater than 0.5, plotted as \emph{dotted} horizontal line.
These differences were also confirmed when we do the $KS$ statistics.

 Interestingly, \citet{Rao2006ApJ...636..610R} have shown that the
 success rate of identifying DLAs can be enhanced if one puts
 additional constraints based on other metal lines, e.g., $R > 0.5$
 and $W$(\mgi) $> 0.1$~\AA. In recent efforts to detect cold gas in
 strong \mgii systems, by using \hi 21-cm absorption, it has been
 found that the detection rate of \hi\ 21-cm absorption is about four
 times higher in systems with strong \feii at $0.5 < z < 1.5$
 \citep[see,][]{Gupta2012A&A...544A..21G, Dutta2017MNRAS.465.4249D}.
 For the average \ew\ (i.e. $\sim 2$~\AA) seen in our sample, the
 probability of \mgii system being DLA is found to be $\sim 50$\%
 \citet[][see their figure 2]{Rao2017arXiv170401634R}. In addition,
 the \nhi\ versus \ew\ relation by \citet{Menard2009MNRAS.393..808M}
 show that the GOTOQs belongs to the sub-DLA systems with log~\nhi\ of
 $\sim$20.1, albeit having a large scatter in \nhi\ up to about 3
 orders of magnitude. \emph{Therefore, a good fraction of our systems
   will be DLAs.}

\subsection{Detection probability of nebular emission lines in  strong \mgii systems}
\label{lab:detection_prob}
The detection of the nebular emission from the \mgii systems not only
depends on the emission line flux but also on (i) the flux of the
background quasar, (ii) the dust attenuation of emission-lines in the
host galaxy and (iii) the bias due to fibre size effects, as discussed
above. Here, we compute the fraction of \oii\ nebular emission line
detections as a function of \ew\ in the SDSS fibre spectra centered
around distant quasar which are detected above the luminosity
threshold of 0.3, 0.5 and 0.7 \lsoii. For this, we restrict ourselves
to a common redshift range of $0.36 \le z \le 0.8$ in SDSS-DR7 and
DR12 and consider the log~\lsoii\ ($\rm erg~s^{-1}$) = 41.6 at the
average redshift of $z = 0.6$ \citep{Comparat2016MNRAS.461.1076C}.
First, we find the number of sightlines suitable for detecting the
\oii\ line with a given luminosity threshold at $\ge 4\sigma$ level by
integrating the error spectra across the expected location of
\oii\ doublet, typically comprises of $\sim$ 12 pixels. Further, the
fraction of systems is computed as a ratio between the number of \mgii
systems with \oii\ emission above a given luminosity threshold to the
total number of \mgii systems for which such a line is detectable.
\par

 In Fig.~\ref{fig:fraction} we show the detection probability of
 \oii\ emission line as a function of \ew\ for the luminosity
 threshold of 0.3 (\emph{top panel}), 0.5 (\emph{middle panel}) and
 0.7 times (\emph{bottom panel}) \lsoii. It is apparent from the
 figure that the detection probability of nebular emission lines
 increases with the strength of the \ew. The fraction of \mgii systems
 detected in emission for various luminosity thresholds (in column 1)
 are listed in the column 2 and 3 of Table~\ref{tab:fraction}. It is
 also clear from the table that the fractional detection of \oii\ in
 \mgii systems is higher for the low luminosity threshold. Last column in
 Table~\ref{tab:fraction}, gives the expected detection rate in the
 case of SDSS-DR12 if one scales the detection rate in SDSS-DR7 by the
 ratio of projected fibre areas. This is much higher than the actual
 rate we find for SDSS-DR12. The difference can be reduced if we
 consider the nebular line emitting regions to be extended and the
 fibre size does affects the observed luminosity as we discussed in
 Section~\ref{lab:fibre_effect}.

Furthermore, in Fig.~\ref{fig:fraction_vs_z} we show the detection
probability of nebular emission from strong (\ew $\ge 1$ \AA) \mgii
systems as a function of redshift for a luminosity threshold of 0.3
\lsoii. A rise in the detection probability of nebular emission as a
function of redshift is clearly seen. This is expected from the
increasing projected area of fibre with $z$ which leads to high
probability for the line emitting regions to come inside the fibre.
Interestingly, in an effort to model the effect of finite fibre size
\citet{Lopez2012MNRAS.419.3553L} have shown that the fraction of
systems for which the absorbing galaxy will give rise to
\oii\ emission in QSO spectra increases as a function of \ew\ and will
depend on the fibre size.  They showed that the detection fraction
  increases from $\sim$50\% for \ew $> 1$~\AA\ to $\sim$90\% for
  \ew\ $> 3$~\AA\ when no limiting flux condition is applied. This is
  what one expects based on the known \ew\ versus $\rho$
  anti-correlation. While what we observe is consistent with the trend
  presented by \citet{Lopez2012MNRAS.419.3553L} the actual detection
  fraction we find is much less than their model prediction because of
  the high luminosity threshold (i.e., 0.3\lsoii) imposed in our
  study. Here, we also find, for a given luminosity threshold, the
\oii\ detection rate in \mgii systems is higher in the SDSS-DR7
spectra than in the SDSS-DR12 spectra (see Fig.~\ref{fig:fraction}).

\begin{figure}
 \epsfig{figure=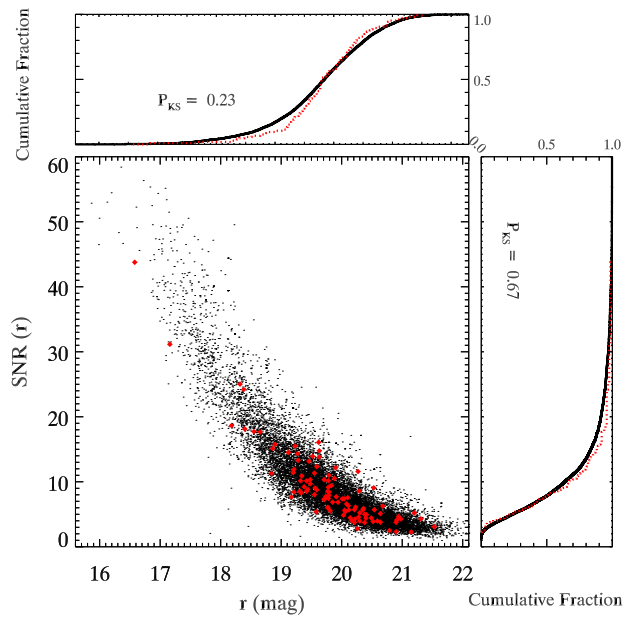,height=8.2cm,width=8.2cm}
\caption{$r$-band $SNR$ versus magnitudes for quasars with strong (\ew
  $> 1$~\AA) intervening \mgii systems (black dots) and systems
  detected in \oii\ nebular emission lines (red diamond) from
  SDSS-DR12. The top and right-hand panels show the cumulative
  distributions of magnitudes and $SNR$, respectively. }
\label{fig:snr_vs_imag}
 \end{figure}

 \begin{table}
 \centering
% \begin{minipage}{120mm}
 {\scriptsize
 \caption{Detected fraction of \mgii absorbers, with \ew\ $\ge 2$~\AA,
   for various luminosity thresholds.}
 \label{tab:fraction}
 \begin{tabular}{@{}  c  c c c @{}}
 \hline  \hline 
 \multicolumn{1}{c}{Criteria}   &\multicolumn{1}{c}{SDSS-DR7}  &       \multicolumn{1}{c}{SDSS-DR12}  &  \multicolumn{1}{c}{expected fraction\textcolor{blue}{}} \\ 
\hline
\\
$\ge 0.3$ \lsoii                       & $0.031^{+0.017}_{-0.012}$      &$0.007^{+0.003}_{-0.002}$     & $0.014^{+0.008}_{-0.005}$  \\
$\ge 0.5$ \lsoii                       & $0.027^{+0.010}_{-0.007}$      &$0.002^{+0.001}_{-0.001}$     & $0.012^{+0.002}_{-0.003}$ \\
$\ge 0.7$ \lsoii                       & $0.011^{+0.017}_{-0.004}$      &$0.001^{+0.003}_{-0.001}$     & $0.005^{+0.008}_{-0.002}$ \\
 \hline                                                                            
 \end{tabular} 
 }
Here, we use  \ls $\rm = 4.16 \times 10^{41}\  erg\ s^{-1}$.
% \end{minipage}
 \end{table}

In Fig.~\ref{fig:snr_vs_imag}, we compare the median $SNR$ measured
over all pixels in the $r$-band to the QSO $r$-band magnitude for all
the QSO spectra from SDSS-DR12 searched here for the nebular
emission and hosting a \mgii system with \ew $\ge 1$~\AA. A $KS-test$
does not show any statistical difference between the $r$-band
magnitude distributions of the two populations with $P_{KS} = 0.23$.
In addition, no difference is seen between the $SNR$ of the two
population with $P_{KS} = 0.67$. Thus it appears that that the quasar
brightness does not impact the detectability of \oii\ nebular emission
for systems with \ew $> 1$~\AA\ \citep[see
  also,][]{Noterdaeme2010MNRAS.403..906N}.

\subsection{The \oiiab\ luminosity:}
{\it Results presented in the last two sections suggest, (1) our
  measured \oii\ nebular line luminosity is an underestimation of true
  luminosity, (2) this bias depends on redshift, and (3) the
  probability of \oii\ detection in \mgii systems increases with
  \ew\ and $z$. Keeping these in mind in what follows we study
  different properties of our \mgii systems with \oii\ nebular
  emission.}

In Fig.~\ref{fig:oiilf} we compare the distribution of measured
\oii\ luminosity of \mgii absorbers detected in SDSS-DR7 and SDSS-DR12
spectra with the \oii\ luminosity functions of galaxies at $z=0.65$
from \citet{Comparat2016MNRAS.461.1076C}. Note that, our measured
luminosities could very well be lower limits as we do not apply any
correction for the dust reddening and the emission line fluxes are
affected by fibre losses. Our measured \oii\ nebular line luminosities
are in the range of $0.14-3.5$~\lsoii\ and $0.09-3.5$~\lsoii\ in case
of SDSS-DR7 and SDSS-DR12 respectively. Therefore, the total
luminosity of these \mgii absorbers are typically higher than
$0.1$\lsoii.

Based on the \oii\ luminosity we derive the SFR using prescription
given by \citet{Kennicutt1998ApJ...498..541K} :
\begin{equation}
  SFR_{[\rm O~II]} = (1.4 \pm 0.4) \times 10^{-41} L_{[\rm O~II]}.
\end{equation}

\noindent In Fig.~\ref{fig:ew_loii} we plot \loii\ versus \ew\ and
show the SFR for individual absorbers in the left side ordinate.
\citet{Bouche2007ApJ...669L...5B} found 67\% of galaxies associated to
the \mgii absorbers with \ew $ > 2$~\AA\ have SFR in the range $1-20~
\rm ~M_{\odot} ~yr ^{-1}$ at $z \sim 2$ based on the H$\alpha$
emission having impact parameter in the range $2-54$ kpc. We note that
about $64$\% \mgii systems in our sample have \ew\ $>2$~\AA\ and SFR
$> \rm 1~M_{\odot} ~yr ^{-1}$. This percentage should be considered as
a lower limit as our flux measurements are affected by fibre size
effects discussed above.

\par

\subsection{\loii\ versus \ew\ correlation ?}
In this section, using our direct detections we investigate the origin
of the strong correlation seen between \ew\ and associated
\oii\ luminosity seen in the stacking analysis of \mgii absorbers
\citep{Noterdaeme2010MNRAS.403..906N,Menard2011MNRAS.417..801M}. It is
important to note that the same fibre size bias affects both these
measurements. In Fig~\ref{fig:ew_loii}, first we compare the \loii\ as
a function of \ew\ from our direct detections. It is clearly visible
that for a given \ew\ the \loii\ has a wide spread. We have performed
the Spearman rank correlation test between \loii\ and \ew\ and found
the correlation coefficient for the \mgii absorbers in SDSS-DR7 fibre
spectrum to be $r_s(\rm SDSS-DR7) = 0.37$ with probability of null
correlation $p_{null}(\rm SDSS-DR7) = 0.003$ (a correlation
significant at $\sim 3\sigma$). We find similar results for our
SDSS-DR12 sample also. Therefore, only a moderately significant
correlation is present between \loii\ and \ew\ when we consider
individual detections. In addition, lack of correlation is confirmed
using survival analysis when we include all the \mgii absorbers from
our sample with \oii\ non-detections as upper limits.

\par

 \begin{figure}
 \epsfig{figure=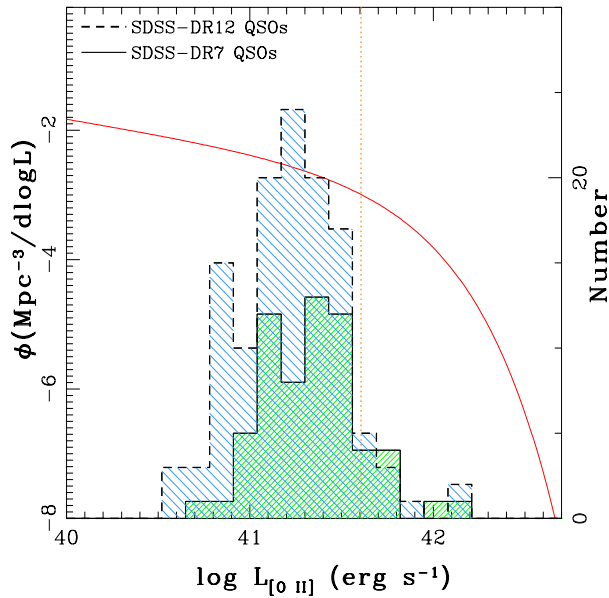,height=8.0cm,width=8.0cm,bbllx=17bp,bblly=144bp,bburx=576bp,bbury=681bp,clip=true}
  \caption{The distribution of \oii\ luminosities (lower limit to be
    precise) for the \mgii absorbers in our sample is compared to the
    field galaxy \oii\ luminosity functions at z = 0.65 (solid curve).
    The vertical dotted line marks the position of \lsoii\ $= 4.1
    \times 10^{41}\ \rm erg\ s^{-1}$
    \citep{Comparat2016MNRAS.461.1076C}.}
\label{fig:oiilf}
 \end{figure}

\begin{figure}
 \epsfig{figure=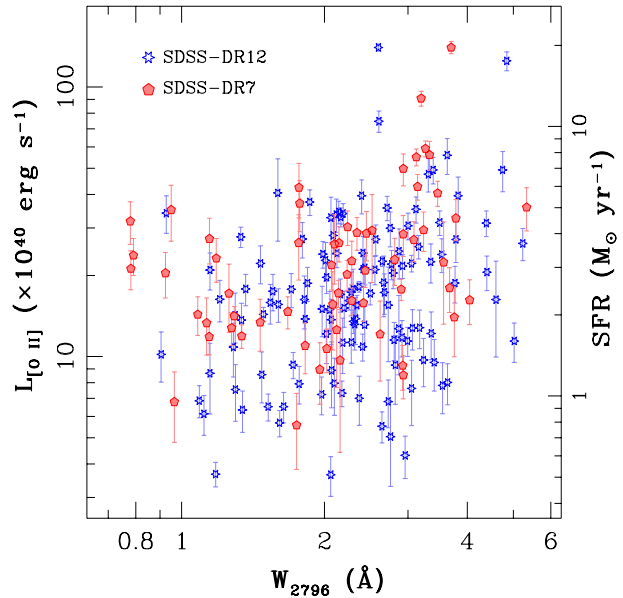,height=8.0cm,width=8.2cm ,bbllx=17bp,bblly=144bp,bburx=592bp,bbury=680bp,clip=true}
\caption{Plot of \oii\ luminosity (\loii) versus  \ew\ for
  the \mgii systems detected the SDSS-DR7(\emph{pentagon}) and
  SDSS-DR12(\emph{stars}). The corresponding star formation rate is
  shown in the right side ordinate.}
\label{fig:ew_loii}
 \end{figure}

\begin{figure*}
 \epsfig{figure=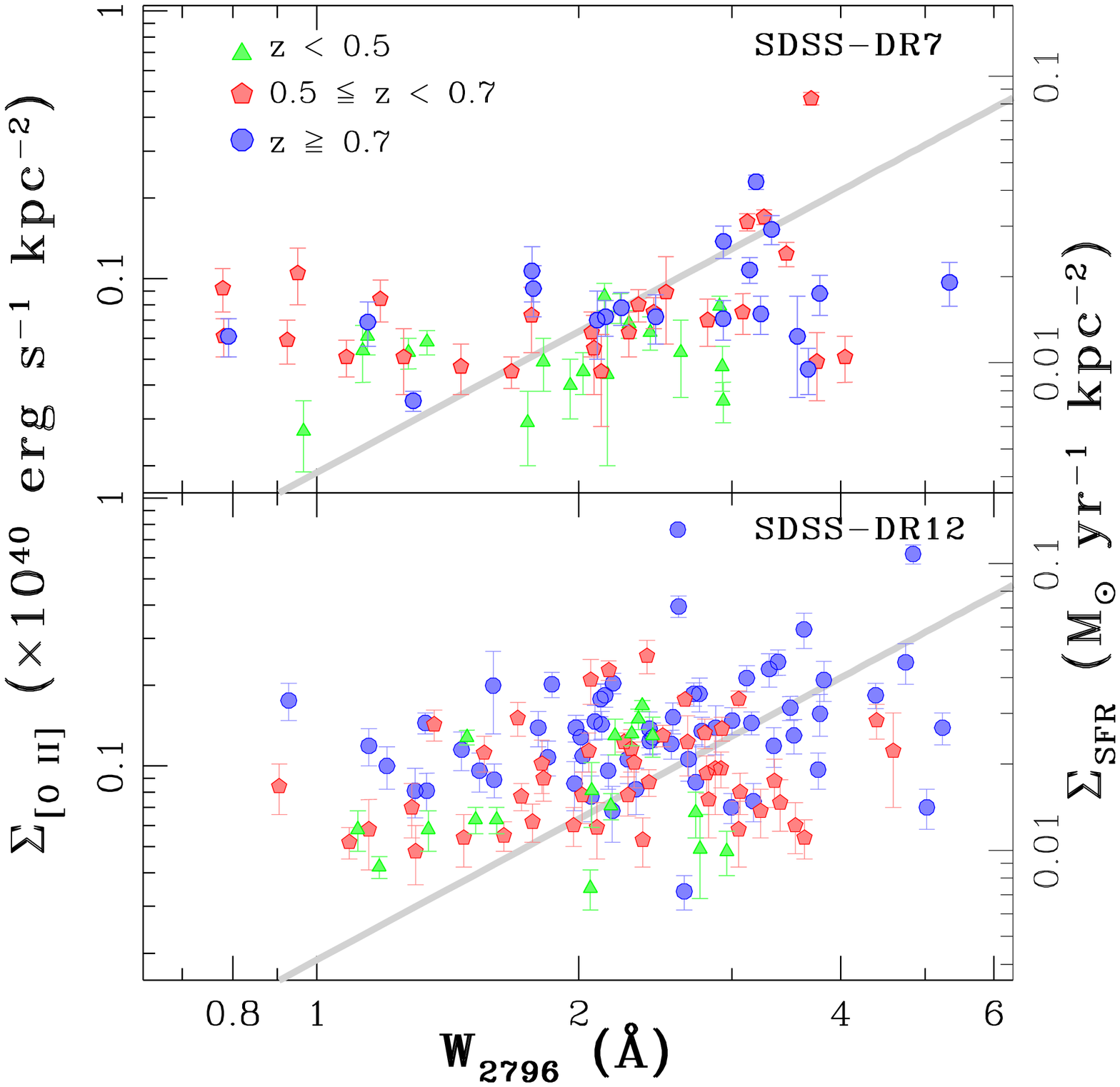,height=8.0cm,width=8.5cm,bbllx=18bp,bblly=144bp,bburx=592bp,bbury=717bp,clip=true}
 \epsfig{figure=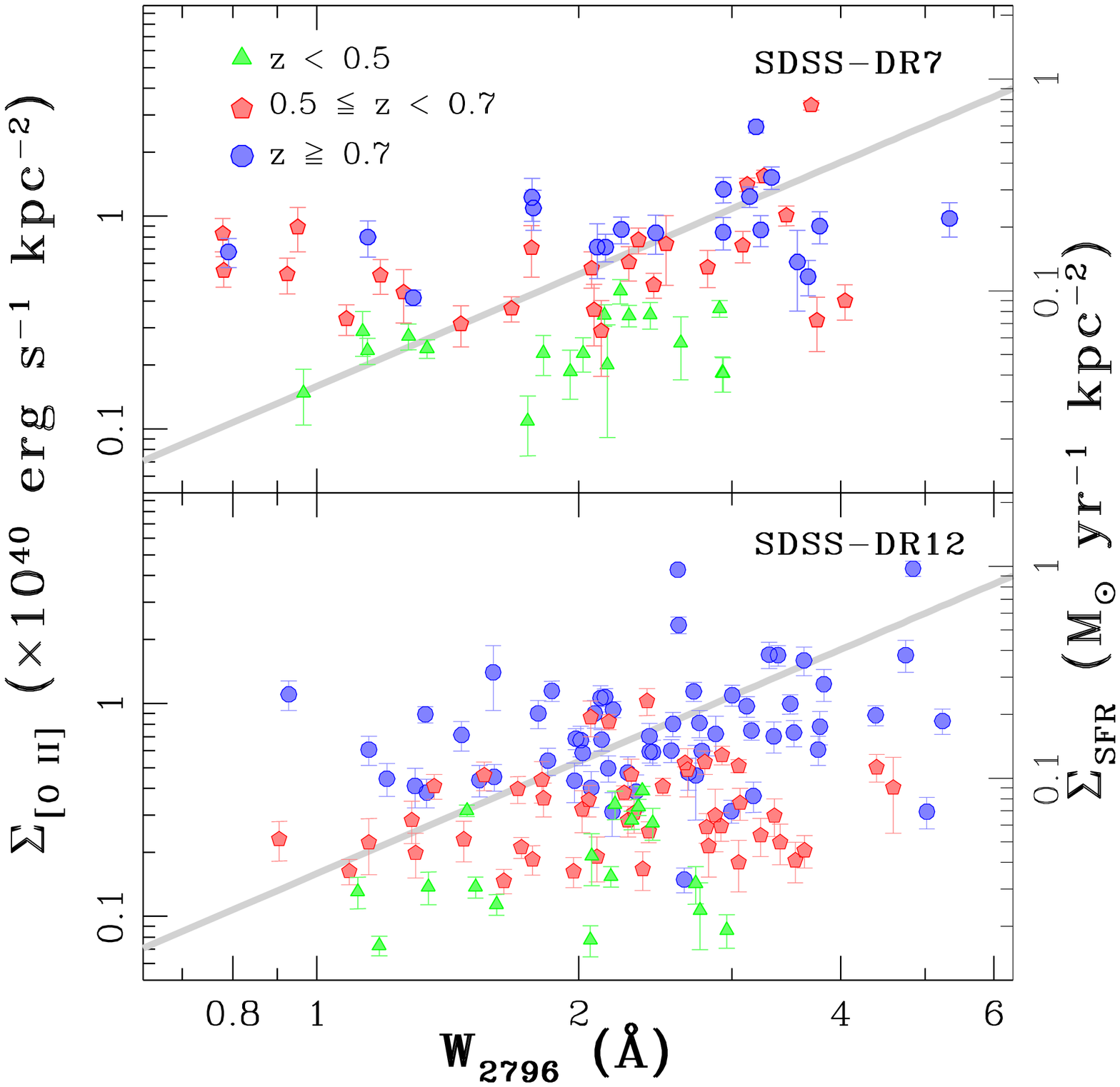,height=8.0cm,width=8.5cm,bbllx=10bp,bblly=144bp,bburx=586bp,bbury=717bp,clip=true}
\caption{ \emph{Left panel:} The \oii\ luminosities surface density
  (\sloii), while considering the surface area equal to the fibre, as
  a function of \ew\ for the \mgii systems detected in the SDSS-DR7
  (\emph{top panel}) and SDSS-DR12 (\emph{bottom panel}),
  respectively. The \emph{solid gray} line shows the best-fit 
  relation between  \sloii-\ew\ obtained from  in the stacking analysis by
  \citet{Menard2011MNRAS.417..801M}. The right side axis show the
  surface star formation rate corresponding to \sloii. \emph{Right
    panel:} The same for the \sloii\ obtained by considering the area
  of a typical galaxy at the absorber redshift.}
\label{fig:sigloii}
 \end{figure*}

Note in \citet{Noterdaeme2010MNRAS.403..906N} and
\citet{Menard2011MNRAS.417..801M} a correlation is found between
luminosity per unit area (\sloii) and \ew. It was argued that the
fibre effects can be taken care of by normalizing the observed line
luminosity by the projected area of the fibre. To compare with this
finding, in Fig.~\ref{fig:sigloii} we show the dependence of surface
brightness of \oii\ emission (denoted by \sloii) as a function of \ew.
For plots in the left panel we assume the surface area to be that of
the fibre and for the plot in the right panel we assume area to be the
average effective area of the galaxies at the corresponding absorber
redshift. In this figure we also show the relationship found by
\citet{Menard2011MNRAS.417..801M}. The surface star formation rate
(\ssfr) derived from \sloii\ are also indicated in the right ordinates
of both the panels. It is clear that our direct detections do not
follow the trend seen in the stacked spectra. A simple correlation
analysis suggests that the correlation could at best be at the level
of $\sim 2 \sigma$. A similar dependence is seen when we compare the
\ew\ with \sloii, obtained by considering the area of the galaxies at
the corresponding absorber redshift. As discussed in
Section~\ref{lab:fibre_effect}, the observed luminosity can be
affected by the fibre size effects that are also redshift dependent.

To see if the above found mild correlation is driven by \ew\ or not,
we divide the sample into three redshift bins of $0.35 \le z < 0.5$
and $0.5 \le z < 0.7$, and $z \ge$ 0.7. It is clear from the figure
that for a given \ew, systems at high redshift show the higher \sloii.
 However, no clear trend is evident in the individual redshift
  bins between \sloii\ and \ew\ in both SDSS-DR7 and SDSS-DR12
  samples.

\emph { In summary, we conclude that among direct detections there is
  no statistically significant correlation between \ew\ and
  \loii\ when we restrict ourselves to small redshift intervals.} This
conclusion is valid for the \sloii\ calculated under both the
assumptions mentioned above. From the discussion presented in
Section~\ref{lab:detection_prob} we find that the fraction of \mgii
absorber showing such nebular emission increases with increasing \ew.
\emph{Therefore, more systems with weaker or no nebular emission have
  contributed to the stacking at low \ew\ compare to those at higher
  \ew.} This could explain the strong correlation found between
\loii\ and \ew\ in the stacked spectra.

\subsection{\loii\ versus $z$}
\label{lab:loiivsz}
We now explore the redshift dependence of \loii\ associated with our \mgii
absorbers. In panel (a) of Fig.~\ref{fig:loii_z} we see a clear
increasing trend of \loii\ with \zabs. The Spearman rank correlations
test finds a strong correlation, with $r_s(\rm SDSS-DR7) = 0.72$ and
$r_s(\rm SDSS-DR12) = 0.74$ between \loii\ and \zabs\ with a null
probability of $p_{null} = 10^{-11}$ (significant at 5.7$\sigma$) and
$\sim 10^{-22}$ (significant at 8.1 $\sigma$), respectively. This
could be a real redshift evolution of the \oii\ luminosity; or some
observational artifact due to (i) luminosity bias for a given flux
threshold as a function of $z$ (as discussed in
Section~\ref{lab:fibre_effect}); (ii) the constant fibre size
corresponding to more projected area at high $z$. We explore these one
by one. \par

    In Panel (a) of Fig.~\ref{fig:loii_z}, the dashed line shows the
    expected luminosity for the observed flux of $10^{-17}\ \rm
    erg\ s^{-1}\ cm^{-2}$, corresponding to the minimum luminosity
    seen in our sample at $z = 0.4$. This line provides a nice lower
    envelope to the observed luminosity at $z < 0.7$.} In the same
       panel the \emph{dot-dashed} lines show the expected luminosity
       of 0.15 and 0.7\lsoii\ as a function of $z$ using the redshift
       evolution of field galaxies luminosity function of \citet[][see
         their Table 7]{Comparat2016MNRAS.461.1076C}. While this
       luminosity range encompasses the observe luminosity at $z <
       0.6$ (barring one system), at high $z$ there are much more
       galaxies brighter than the 0.7\lsoii\ galaxies. As discussed in
       Section~\ref{lab:fibre_effect}, lower the redshift higher will
       be reduction in the measured luminosity compared to the actual
       luminosity due to fibre effects. Thus, the dominant factor for
       this \loii-$z$ relationship explored in the plots could be the
       redshift dependence of the fiber losses. In addition, the lack
       of low luminosity detections at high-$z$ is probably biased due
       to the observing strategy in SDSS-DR12 where to maximize the
       flux in the blue part an offset was applied to the position of
       the quasar target fibres to compensate for atmospheric
       refraction \citep[see also][]{Paris2012A&A...548A..66P}.

We can account for the relative increase in characteristic luminosity
of galaxies as a function of $z$ by scaling down the observed
\loii\ by a factor $f =$ \lsoii($z=0$)/\lsoii($z$)
\citep{Comparat2016MNRAS.461.1076C}. The scaled luminosity is shown in
the panel (b) of Fig.~\ref{fig:loii_z}. The Spearman rank correlation
test finds a correlation with $r_s(\rm SDSS-DR7) = 0.48$ and $r_s(\rm
SDSS-DR12) = 0.45$ between \loii\ and \zabs\ with a null probability
of $p_{null} = 10^{-3}$ (significant at 3.6$\sigma$) and $\sim
10^{-8}$ (significant at 5.3$\sigma$), respectively. This indicates
that the above correlation is not dominated mainly by the redshift
evolution of luminosity. This once again confirms that the fibre size
effect is a dominant effect.

\par

\begin{figure*}
 \epsfig{figure=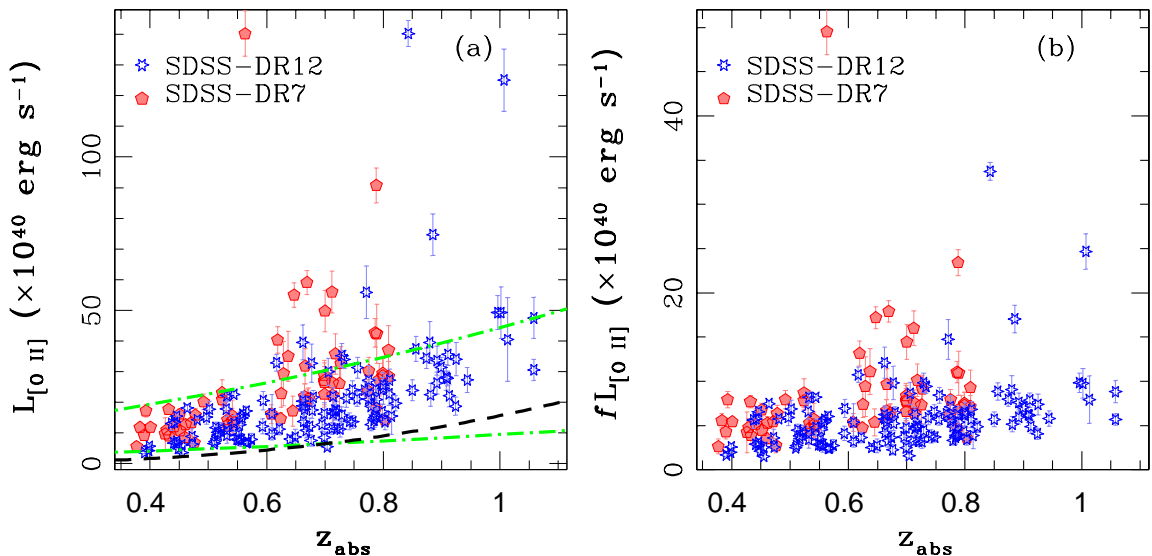,height=7.5cm,width=15.2cm,bbllx=30bp,bblly=425bp,bburx=568bp,bbury=691bp,clip=true}
\caption{ \emph{Panel a:} The \oii\ luminosity (\loii) of \mgii
  absorbers with nebular emission as a function of redshift within 3
  arcsec fibre (\emph{pentagon}) in the SDSS-DR7 and within 2 arcsec
  fibre (\emph{stars}) in SDSS-DR12 spectra. The \emph{lower
    dot-dashed line} shows the expected \oii\ luminosity from a
  0.15\lsoii\ galaxy as a function of redshift. This provides a lower
  envelop to the observed data. The \emph{upper dot-dashed line} shows
  the expected \oii\ luminosity from a 0.7\lsoii\ galaxy as a function
  of redshift. \emph{Panel b:} The scaled \loii, after correcting for
  the evolution of \lsoii\ with $z$, where
  $f=$\lsoii($z=0$)/\lsoii($z$). }
\label{fig:loii_z}
 \end{figure*}

\emph{In summary, as shown by the field galaxies, \oii\ luminosity of
  our \mgii absorbers are higher at higher $z$. The correlation
  remains even after we account for redshift evolution of \loii.
  Therefore, we conclude that the strong correlation seen between
  \loii\ and $z$ seen in our sample (as well as stacked spectra in the
  literature) is influenced substantially by the redshift dependent
  fibre losses.}

\subsection{ {[O~{\sc iii}]/[O~{\sc ii}]} and \ohb\ nebular line ratio :}
\label{sub:o3o2}

The \o3o2 and \ohb\ ratios are sensitive to the hardness of the
ionizing radiation field, and serves as a ionization parameter
diagnostic of a galaxy \citep{Baldwin1981PASP...93....5B,
  Kewley2001ApJ...556..121K}. A rise in ratios of nebular emission
lines, in particular \o3o2 and \ohb, of star-forming galaxies is seen
over the redshift range
$z=0-5$~\citep[see][]{Nakajima2014MNRAS.442..900N,
  Steidel2014ApJ...795..165S,Kewley2015ApJ...812L..20K,
  Khostovan2016MNRAS.463.2363K}. This implies that typical galaxies at
high redshifts have higher ionization parameter, lower metallicity,
harder stellar ionizing radiation field and higher electron densities
compare to those of local galaxies. The direct detection of nebular
emission lines from \mgii absorbers allows us to explore their
physical conditions. We assume that the nebular line ratios do not
depend on the fibre size used in the SDSS-DR7 and SDSS-DR12 spectra.
Therefore, while studying the emission line ratios we have combined
the SDSS-DR7 and SDSS-DR12 sample of \mgii absorbers with nebular
emission.

\begin{table*}
 \centering
 \begin{minipage}{120mm}
 {\small
 \caption{Redshift evolution of line ratios, metallicity and
   ionization parameter of \mgii absorbers.}
 \label{tab:Z}
 \begin{tabular}{@{} c c c c c c  @{}}
 \hline  \hline 
 \multicolumn{1}{c}{redshift($z$)}   &\multicolumn{1}{c}{systems}  &   \multicolumn{1}{c}{log~(\o3o2)}  &  \multicolumn{1}{c}{log~(\ohb)} &  \multicolumn{1}{c}{log ($Z$)}  &  \multicolumn{1}{c}{log ($q$) }\\
\hline \\

$0.10 \le z < 0.35$ & 79   &  $-0.13\pm 0.03$ &  $0.35\pm 0.04$    &   $8.33^{+0.19}_{-0.19} $  &    $7.52^{+0.16}_{-0.10}$             \\
$0.35 \le z < 0.65$ & 34   &  $-0.02\pm 0.03$ &  $0.44\pm 0.04$    &   $8.33^{+0.13}_{-0.19} $  &    $7.62^{+0.13}_{-0.10}$             \\
$0.65 \le z < 1.10$ & 42   &  $+0.08\pm 0.03$ &  $0.62\pm 0.05$    &   $8.30^{+0.10}_{-0.10} $  &    $7.69^{+0.10}_{-0.10}$             \\
 
 \hline                                                                                 
 \end{tabular} 
 }                                                   
 \end{minipage}
 \end{table*}

\begin{figure*}
 \epsfig{figure=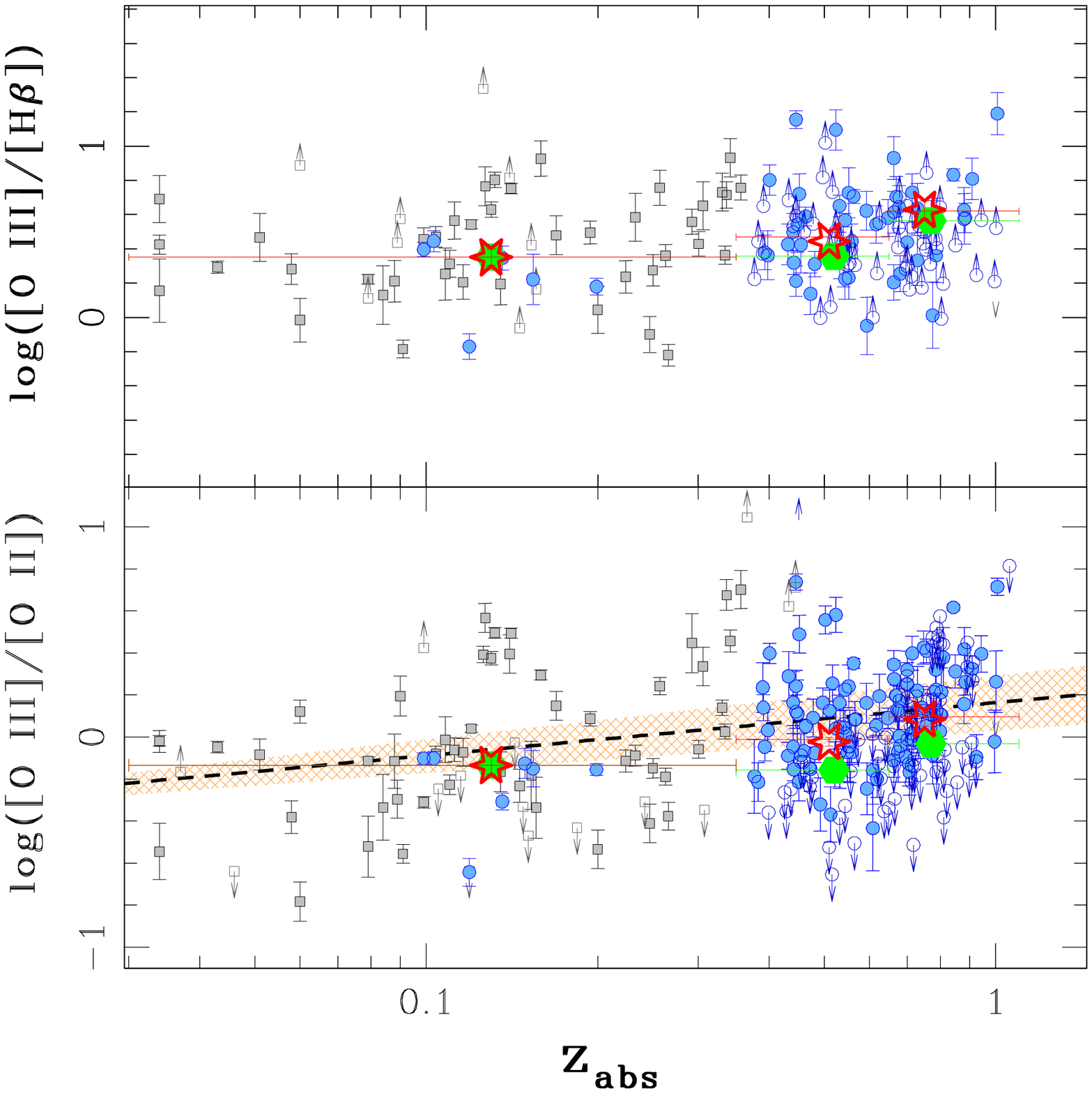,height=8.2cm,width=8.2cm}
 \epsfig{figure=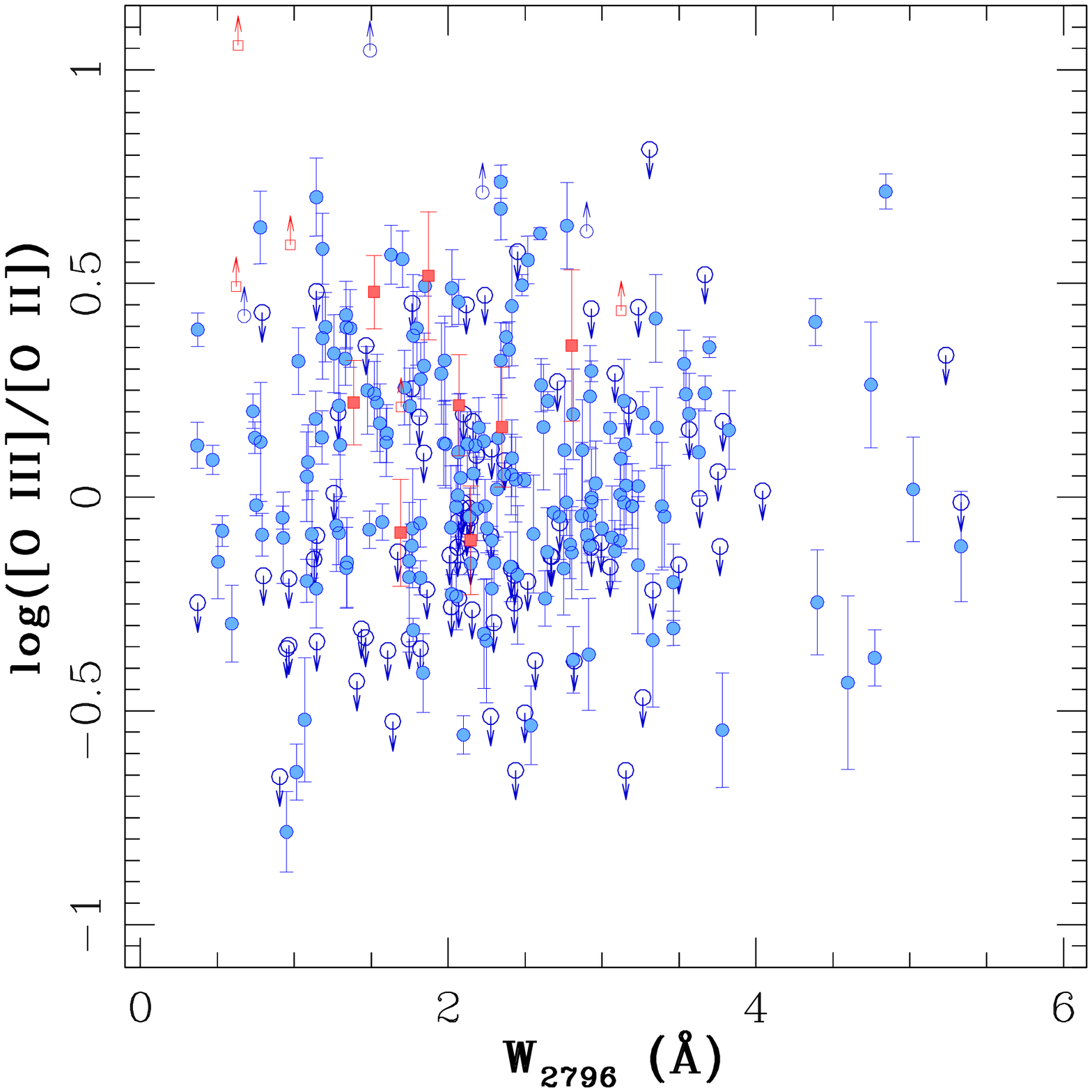,height=8.2cm,width=8.2cm}
\caption{ \emph{Left panel:} The [O~{\sc iii}]/H$\beta$ (\emph{top})
  and \o3o2 (\emph{bottom}) line ratio as a function of $z$. The \mgii
  systems detected in nebular emission line in our sample are shown in
  \emph{circles} while the systems detected by
  \citet{Straka2015MNRAS.447.3856S} are shown in \emph{squares}. The
  measured \o3o2 ratio in the stacked spectra over three redshift bins
  is shown in \emph{hexagon}. The ratio while considering only the
  systems with \oiii\ detected at $\ge 3\sigma $ is shown as
  \emph{filled circles/squares}. The upward(downward) arrows represents the
  $3\sigma$ upper(lower) limits. The redshift evolution of \o3o2, best
  described by a power-law fit by \citet{Khostovan2016MNRAS.463.2363K}
  is shown as \emph{dashed line} along with a 1$\sigma$ uncertainty
  (\emph{shaded region}). \emph{Right panel:} The \o3o2 ratio as
  function of \ew. The \emph{squares} represent the systems with
  strong \oiii\ emission.}
\label{fig:o3o2}
 \end{figure*}

First, we explore the dependence of \o3o2 and \ohb\ nebular line
ratios with redshift (see Fig.~\ref{fig:o3o2}). While computing the
line ratios for cases where the \oiii\ or \hbeta\ line is not detected
we use the $3 \sigma$ upper limits. In addition, we have excluded
systems for which both the lines are not detected. In
Fig.~\ref{fig:o3o2} the limits are shown as (\emph{arrows}) and the
detections ($\ge 3 \sigma$) are shown with \emph{circles}. For a
subset of galaxies with firm detections we detect a significant
correlation between the \o3o2 line ratio with $z$, with a Kendall's
rank correlation coefficient ($r_k$) of 0.3 with null probability
$p_{null} = 0.006$ which is significant at $\sim 2.7 \sigma$ level
assuming Gaussian statistics. A similar correlation is seen for our
entire sample with $r_k$=0.2 and null probability $p_{null} = 0.0007$
where we include the upper limits as censored data points and perform
survival analysis using the {\sc `cenken'} function in the {\sc
  `nada'} package of {\sc r} . This suggests that the of redshift
evolution of nebular line ratio in \oii\ detected \mgii systems
follows the trend shown by the field galaxies. However, we do not find
any correlation between \ohb\ line ratio and $z$ with $r_k = -0.008,
-0.02, $ and $p_{null}= 0.87, 0.77$ for the entire sample as well for
the clear detections.

Next, we compute the average line ratio using the composite spectrum
obtained for the systems with \oii\ detections in SDSS-DR12. To
generate the composite spectrum, an individual spectrum is shifted to
the rest-frame of the \mgii absorber while conserving the flux and
rebinning on to a same logarithmic scale of a pixel to wavelength used
in the SDSS \citep{Bolton2012AJ....144..144B}. Each spectrum is
subtracted with the best-fit principle component analysis (PCA)
continuum model of \citet{Bolton2012AJ....144..144B} and the residual
spectra is combined together using a median statistics \citep[see
  also,][]{Joshi2017MNRAS.465..701J}. Note that, for most of the \mgii
systems detected in emission, with redshift range of $0.4-1.1$, the
\oiii\ and \hbeta lines fall in the region affected by poorly
subtracted sky emission lines. Therefore, to avoid any contamination
from sky residuals we have generated two sets of composite spectra by
considering the entire sample with upper limits and another with
excluding the systems with upper limits for \oiii. We divide our
sample into two redshift bins of $0.36-0.65$ and $0.65-1.10$, consists
of about 49 and 74 galaxies per bin in the composite spectra including
upper limits. The number of galaxies over the above $z$-bins reduces
to 34 and 42 respectively if we consider the systems with only clear
detections. \par

The nebular emission line ratio for \o3o2\ and \ohb\ in a stacked
spectrum of the systems with clear detections, over various redshift
bins are listed in column 3 and 4 of Table~\ref{tab:Z}. In the stacked
spectra of our entire sample the log (\o3o2) is found to be evolving
from $-0.10 \pm 0.02$ to $-0.03 \pm 0.02$, by 0.07 dex over the
redshift range of $0.36 \le z \le 1.0$. In addition, a clear
increasing trend of \o3o2 with $z$ is apparent in the stacked spectra
where all the emission lines are clearly detected (see
Fig.~\ref{fig:o3o2}, \emph{lower left panel}). We find a change of
$\sim$0.06 dex in \o3o2 which increases from $0.95 \pm 0.06$ to $1.22
\pm 0.09$ between the median redshift of $z \sim$0.52 and $\sim$0.76.
Using the star-forming galaxies from High-$z$ Emission Line Survey
(HiZELS), \citet{Khostovan2016MNRAS.463.2363K} have shown that the
\o3o2 evolves out to $z$ $\sim$5 and is best described by a power law
of the form, \o3o2 $= (0.59\pm 0.07)\times (1+z)^{(1.17 \pm 0.24)}$
(see their eq. 6). It also shows a similar change of $\sim$0.07 dex in
\o3o2\ line ratio for the above redshifts. In addition, we measure the
\o3o2\ and \ohb\ ratio in a composite spectrum of low-$z$ (i.e., $z <
0.35$) galaxies, detected in the nebular emission lines in fibre
spectra of a background quasars without a prior knowledge of the
absorption \citep[see e.g.,][]{Straka2015MNRAS.447.3856S}. It is clear
from Fig.~\ref{fig:o3o2} (\emph{left lower panel}) that \o3o2 rises by
$\sim$0.21 dex between $0.1\le z \le 1.1$. Similarly, the \ohb\ is
also found to be evolving from $2.25\pm0.20$ to $4.25\pm 0.47$ by
$\sim$0.27 dex between $0.1\le z \le 1.1$ (see Table~\ref{tab:Z}).
\emph{The above trends indicate that the nebular emissions seen in
  \mgii systems in our sample follow the trends shown by the general
  population of normal star-forming galaxies.}

 In a stacking analysis of galaxies between $0.2 < z < 0.6$ from the
 SHELS galaxy redshift survey, \citet{Kewley2015ApJ...812L..20K} have
 studied the \o3o2\ and \ohb\ line ratio as a function of stellar mass
 bins, derived from the broad-band photometry. They showed that the
 optical line ratio is also a strong function of stellar mass, where
 the galaxies with lower stellar masses show higher line ratio
 \citep[see also,][]{Henry2013ApJ...769..148H, Ly2014ApJ...780..122L,
   Hayashi2015PASJ...67...80H,Ly2015ApJ...805...45L}. The \o3o2\ and
 \ohb\ in our \mgii systems are found to be similar as seen in the
 field galaxies with stellar masses in the range $9.2 < \rm
 log(M/M_{\odot}) < 9.4$ \citep[][see their figure 2 and
   3]{Kewley2015ApJ...812L..20K}.

\begin{figure*}
  \epsfig{figure=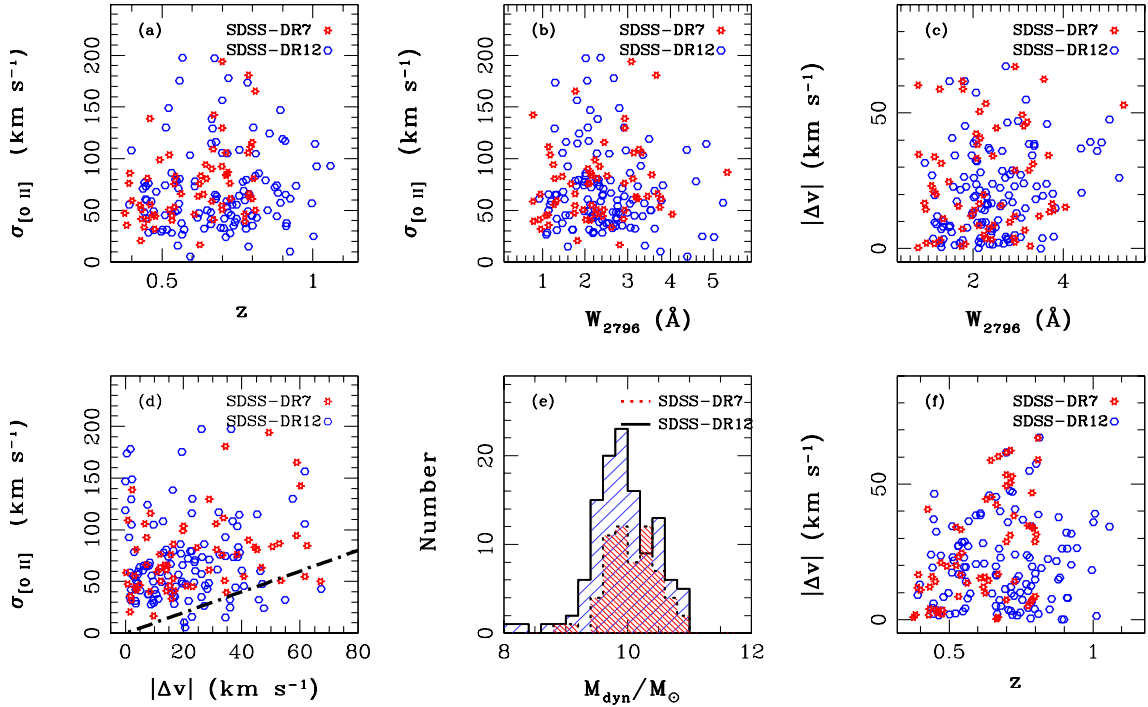,height=9.5cm,width=15.0cm,bbllx=34bp,bblly=345bp,bburx=565bp,bbury=690bp,clip=true}
\caption{Comparison of the kinematic properties of \mgii absorbers
  detected in SDSS-DR7(\emph{stars}) and SDSS-DR12(\emph{hexagon})
  spectra. Panels a and b show the dependence of velocity widths
  (\sigoii) on redshift and the \mgii rest equivalent width (\ew).
  Panel c and f show the dependence of the absolute velocity offset
  ($|\Delta v|$) between the absorption and emission-line redshift
  with \ew\ and redshift. Panel d shows the \sigoii\ as a function of
  $|\Delta v|$. The \emph{dot-dashed line} represents the \sigoii $ =
  |\Delta v|$ line. Panel e shows the distribution of the dynamical
  mass of \mgii absorbers detected in SDSS-DR7 (\emph{dotted
    histogram}) and SDSS-DR12 (\emph{solid histogram}) spectra.}
\label{fig:delv}
 \end{figure*}

 We measure the gas phase metallicity (Z) and ionization parameter (q)
 of the ionized nebula using the {\sc~izi} (Inferring metallicity and
 ionization parameters) code described in
 \citet{Blanc2015ApJ...798...99B} and assuming the photoionization
 model results of \citet{Levesque2010AJ....139..712L}. The Z and q for
 the three $z$ bins are listed in column 5 and 6 of Table~\ref{tab:Z}.
 A typical metallicity of \mgii systems over a redshift range of $0.1
 \le z \le 1.1$ is found to be sub-solar with log Z = 8.3. \par

In Fig.~\ref{fig:o3o2} (\emph{right panel}), we plot the \o3o2 as a
function of \ew. We find that the \o3o2 does not depend on the \ew,
with Kendall's rank correlation coefficient ($r_k$) of 0.02 and null
probability $p_{null} = 0.62$ in a survival analysis while including
the upper limits as censored data points. In addition, no correlation
is seen for systems with clear detection with $r_k = -0.07$ and
$p_{null}=0.22$.

\subsection{Velocity width of emission and mass of \mgii absorbers}

In this section, we study the kinematic properties of \mgii systems in
our sample. The velocity widths (deconvolved for the instrumental
broadening) of \oiiab\ doublet (i.e., \sigoii) is found in the range
of $\sim 17\pm16$~\kms\ to $194\pm 72$~\kms\ in SDSS-DR7 and $6 \pm
11$~\kms\ to $198\pm 91$~\kms\, in the SDSS-DR12 spectra with an
average \sigoii\ of $\sim$ 75~\kms. This is similar to a typical
velocity dispersion found in the SDSS galaxies
\citep{Oh2011ApJS..195...13O}. The $KS-test$ does not indicate any
differences between the \sigoii\ in SDSS-DR7 and DR12 with $p_{null} =
0.23$. We also measure the velocity shift ($|\Delta v|$) between the
absorption and emission redshift for each system. These values are
found to be typically $< 70$~\kms. Now we compare the \sigoii\ and
($|\Delta v|$) with the other parameters of the systems.

In panel (a) of Fig.~\ref{fig:delv} we look for the dependence of
\sigoii\ as a function of redshift of the \mgii absorbers. For the
systems detected in SDSS-DR7 fibre spectra a Spearman rank correlation
test finds a correlation with $r_s$(SDSS-DR7) = 0.41 and null
probability $p_{null} = 0.0008$ (significant at $\sim 3.2\sigma$).
However, no such correlation is seen between \sigoii\ and $z$ for the
systems detected in SDSS-DR12 with $r_s$(SDSS-DR12) = 0.09 and
$p_{null} = 0.3$. For comparison when we use field galaxies from
MPA-JHU SDSS-DR7 catalog we find no correlation between $z$ and
\sigoii\ at $> 1.5\sigma$.

In panel (b) of Fig.~\ref{fig:delv} we compare \ew\ with \sigoii. We do
not find any correlation between \sigoii\ and \ew\ of \mgii absorber
with a correlation coefficient of $r_s$(SDSS-DR7) = $-0.11$ and
$r_s$(SDSS-DR12) = 0.22 at a significance level of $< 2 \sigma$. If
\sigoii\ can be treated as a proxy to the underlying galaxy mass than
the above result suggests that there is no one to one correspondence 
between \ew\ and mass of the host galaxy. \par 

We also compare the relative velocity shift ($|\Delta v|$) between
absorption and emission line as a function of \ew\ in panel (c) of
Fig.~\ref{fig:delv}. The Spearman rank correlation test finds no
correlation at $> 2\sigma$ between \ew\ and $|\Delta v|$ in the
SDSS-DR7 and DR12. The maximum $|\Delta v|$ is found to be $\sim$
70~\kms. Even though there is no correlation present it is interesting
to note that systems with \ew $> 4$~\AA\ have larger $|\Delta v|$.
Further, in panel (f) of Fig.~\ref{fig:delv} we show the dependence of
$|\Delta v|$ on redshift. For the \mgii systems detected in SDSS-DR7
spectra the $|\Delta v|$ shows a correlation with $z$ with
$r_s$(SDSS-DR7) = 0.47 and $p_{null} =0.0001$ (significant at
$3.6\sigma$). However, no such correlation is seen for the \mgii
systems detected in SDSS-DR12 spectra with $r_s$(SDSS-DR12) = 0.05 and
$p_{null} =0.60$.

 In panel (d) of Fig.~\ref{fig:delv}, we compare the \sigoii\ with the
 absolute relative velocity shift ($|\Delta v|$) between absorption
 and emission line. Note that, it is quite possible that the
 absorption and emission may come from two different regions of the
 galaxy. We do not find any correlation between \sigoii\ and $|\Delta
 v|$. It is clear from the figure that all but 14 systems are well
 within a regime where gas is bound to the host galaxy, i.e, $|\Delta
 v| \le$ \sigoii, shown as \emph{dot-dashed line} in panel (d) of
 Fig.~\ref{fig:delv}. We note that among the 14 systems where $|\Delta
 v| >$ \sigoii, 8 systems belongs to the ultrastrong \mgii absorbers,
 i.e., with \ew\ $\ge$ 3~\AA. In addition, 2 systems are having $2.5
 \le$ \ew\ $<$ 3.0~\AA\ and 4 systems with $1 \le$ \ew\ $< 2.5$~\AA.
 As these systems are predominantly ultrastrong \mgii absorbers, they
 may be produced in the outflows from the galaxies
 \citep{Nestor2011MNRAS.412.1559N,Gauthier2013MNRAS.432.1444G}. The
 fraction of ultrastrong \mgii absorbers with $|\Delta v| >$
 \sigoii\ is found to be $\sim 18\%$ (i.e., 8 out of 43 systems).
 However, these galaxies do not show signatures of high SFR. In order
 to draw a firm conclusion on winds we need high resolution spectra
 that will resolve the velocity profiles of absorption line
 \citep{Kacprzak2013ApJ...777L..11K,Ho2017ApJ...835..267H}.

Next, by using the width of \oii\ line we measure the dynamical mass
(\mdyn) of the \mgii absorbers using,

\begin{equation}
M_{\rm dyn} = C \frac{r_{eff} \sigma^2}{G}.
\label{eq:mdyn}
\end{equation}

\noindent Here, C is a geometric correction factor that can vary
depending on the assumed shape and orientation of the galaxies,
$r_{eff}$ is the effective half-light radius, and G is the
gravitational constant. We have used $C =3$ similar to
\citet{Maseda2013ApJ...778L..22M}. The $r_{eff}$ is computed from the
scaling relation of $r_{eff}$ versus $z$ for the star forming galaxies
by \citet[][see above
  Section~\ref{lab:loiivsz}]{Paulino-Afonso2017MNRAS.465.2717P} at the
redshift of \mgii absorber. The distribution of \mdyn\ for the \mgii
systems detected in the SDSS-DR7 (\emph{dotted histogram}) and
SDSS-DR12 (\emph{solid histogram}) are shown in panel (e) of
Fig.~\ref{fig:delv}. The \mgii systems in our sample probe a range of
dynamical mass with log~\mdyn (M$_{\odot}$) = $8.8-11.0$ and
$7.9-10.0$ with a median of $\sim 10.0$ and $\sim 9.9$ in the SDSS-DR7
and SDSS-DR12, respectively. Note that, the above dynamical masses
represent a lower limit on the mass as the measured $\sigma$ could
suffers from finite fibre loss.

\emph{In summary, we find that the \mgii systems with \oii\ emission
  in our sample share similar properties like normal galaxies with
  typical line width of $\sim 75$~\kms\ and dynamical mass of,
  log~\mdyn (M$_{\odot}$) $\sim 9.9$. In addition, most of the
  absorbers seem to be bound to the host galaxy.}

\begin{figure}
 \epsfig{figure=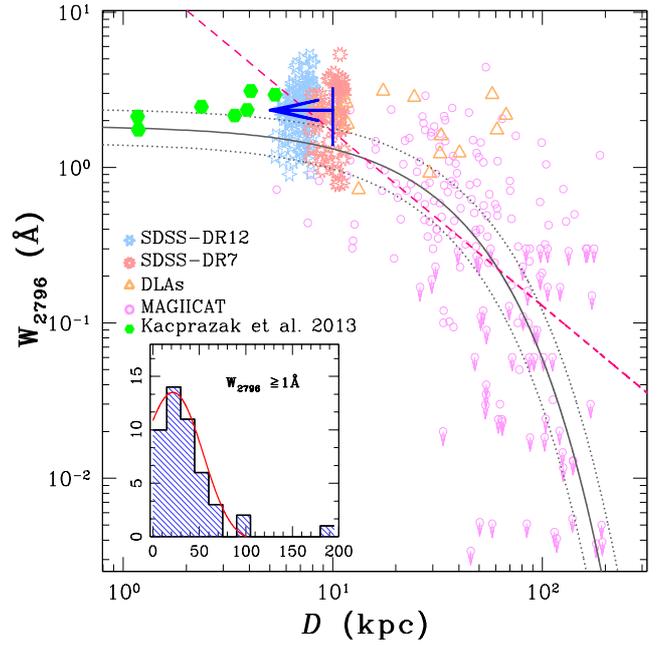,height=8.5cm,width=8.5cm,bbllx=18bp,bblly=167bp,bburx=587bp,bbury=715bp,clip=true}
 \caption{\mgii equivalent width versus impact parameter ($\rho$) for
   the systems detected in nebular emission (\emph{star}). The
   spectroscopically identified \mgii galaxies from MAG{\sc ii}CAT
   survey are presented as open circle, whereas those with upper
   limits on absorption are shown with downward arrows. The
   \emph{hexagons (green)} show the seven \mgii galaxies detected at
   impact parameters of $\le 6 \rm kpc$ by
   \citet{Kacprzak2013ApJ...777L..11K} at $z < 0.1$. The \emph{pink},
   dashed curve is the power law fit obtained by
   \citet{Chen2010ApJ...714.1521C}. The solid curve is a log-linear
   maximum likelihood fit given by \citet{Nielsen2013ApJ...776..115N}
   while dotted curves provide 1$\sigma$ uncertainties in the fit. The
   inset shows the $\rho$ distribution for the \mgii absorbers with
   \ew\ $\ge 1$~\AA. The \emph{solid cure} represents the skewed
   Gaussian fit to the $\rho$ distribution.}
 \label{fig:ewvsd}
 \end{figure}

\subsection{\ew\ versus impact parameter:}
\label{sub:impact}
 It has been firmly established that the rest-frame equivalent width
 of \mgii absorption is anti-correlated with the impact parameter,
 $\rho$ \citep{Bergeron1991A&A...243..344B,Steidel1995qal..conf..139S,
   Chen2010ApJ...714.1521C,Rao2011MNRAS.416.1215R,Nielsen2013ApJ...776..115N,Nielsen2013ApJ...776..114N},
 albeit with a large scatter. Recently, using a sample of 182 galaxies
 having impact parameters of $\rho \ge 10 \rm ~ kpc$,
 \citet{Nielsen2013ApJ...776..114N,Nielsen2013ApJ...776..115N} have
 shown a $7.9\sigma$ anti-correlation between \ew\ and $\rho$ which is
 well represented by a log-linear relation of $\rm log~$\ew = $\rm
 (-0.015\pm 0.002) \times \rho + (0.27 \pm 0.11)$. However, a
 considerable scatter is seen in the data which may be related to the
 galaxy luminosity, where more luminous galaxies have larger \ew\ at a
 fixed $\rho$
 \citep[see,][]{Nielsen2013ApJ...776..115N,Churchill2013ApJ...779...87C}.
 Furthermore, using $\sim$7 spectroscopically confirmed low redshift
 ($z \sim 0.1$) galaxies \citet{Kacprzak2013ApJ...777L..11K} have
 shown that anti-correlation between \ew\ and $\rho$ is maintained
 even at low impact parameter of $\le 6\ \rm kpc$ . \par

 Here, we explore the \ew\ versus $\rho$ correlation at low impact
 parameters over a large redshift range of $0.3 \le$ \zabs\ $\le 1.1$.
 Note that, for the \mgii systems at $0.3 \le$ \zabs\ $\le 1.1$, the
 fibre diameter of 3 and 2 arcsec ensures a close star-forming galaxy
 within an impact parameter of $8.1-12.3\ \rm kpc $ and $5.4-8.2\ \rm
 kpc$, respectively. In Fig~\ref{fig:ewvsd} we show the distribution
 of \ew\ and $\rho$ for our sample in the \ew-$\rho$ plane. The circle
 and upper limits are for the absorbers and non-absorbers from the
 spectroscopically identified galaxies hosting \mgii absorbers
 compiled in the MAG{\sc ii}CAT
 \citep{Nielsen2013ApJ...776..114N,Nielsen2013ApJ...776..115N}. We
 also show a best-fit power law from \citet{Chen2010ApJ...714.1521C},
 as a \emph{dashed} curve. At a fixed impact parameter of $\sim 5\ \rm
 and\ 10\ kpc$ the above log-linear relation by
 \citet{Nielsen2013ApJ...776..114N} predicts a \ew\ of
 1.6$^{+0.49}_{-0.38}$~\AA\ and 1.32$^{+0.46}_{-0.34}$ \AA. Whereas,
 at these impact parameters the power-law fit from
 \citet{Chen2010ApJ...714.1521C} predicts a \ew\ of
 $2.9^{+1.7}_{-1.1}$~\AA\ and $1.3^{+0.9}_{-0.5}$~\AA, respectively.
 The median \ew\ probed in our sample is found to be 2.23 and
 2.32~\AA\ for SDSS-DR7 and SDSS-DR12 data set, respectively. It is
 clear from Fig.~\ref{fig:ewvsd} that majority of the points lie above
 the log-linear relation of \ew\ versus $\rho$. \par

  Next, we estimate the average \ew\ of \mgii systems expected in our
  case by assuming the above two functional forms for \ew-$\rho$
  relation and associating an impact parameter dependent detection
  probability of \mgii systems as:

 \begin{equation}
\rm \langle W_{2796} \rangle = \frac {\int_{0}^{\rho_{max}} W(\rho)
  p(\rho)d\rho} {{\int_{0}}^{\rho_{max}}   p(\rho)d\rho} .
\label{eq:ewvsd}
\end{equation}

\noindent Here, $\rm W(\rho)$ is \ew-$\rho$ relation, $\rm \rho_{max}$
is the maximum impact parameter, taken as a projected fibre radius of
10 kpc at a median redshift of 0.65 for the 3 arcsec SDSS-DR7 fibre.
$\rm p(\rho)$ is impact parameter dependent detection probability of
\mgii absorber which is defined as $\rm p(\rho) = \int_{0}^{\rho} 2
\pi \rho d\rho/ \pi \rho_{max}^2$, assuming a spherical halo. Using
eq.~\ref{eq:ewvsd} and the best fit log-linear and power-law relations
for \ew-$\rho$ we find an average \ew\ up to $\rho$ $\sim$ 10 kpc to
be $1.4^{+0.6}_{-0.4}$~\AA\ and $2.1^{+0.4}_{-0.3}$~\AA, respectively.
However, if we consider the impact parameter corresponding to a 2
arcsec fibre used in SDSS-DR12 at median redshift, i.e., of $\sim$ 7
kpc, the \ew\ for above two models are found to be
$1.6^{+0.5}_{-0.4}$~\AA\ and $3.2^{+0.5}_{-0.4}$~\AA. The median value
found for our sample, while consistent with these average values
within $2\sigma$, is closer to the power-law prediction.

Furthermore, using the $\rho$ distribution of spectroscopically
confirmed strong, \ew\ $\ge 1$~\AA, \mgii absorbers from MAG{\sc
  ii}CAT (i.e. $\sim$ 40 systems) along with the seven systems from
\citet{Kacprzak2013ApJ...777L..11K} we computed the detection
probability of strong \mgii absorbers within impact parameter of $\le
8 \rm ~kpc$ (a typical projected size of SDSS fibre). The distribution
of the $\rho$ is shown in the inset of Fig.~\ref{fig:ewvsd}. We have
modelled the $\rho$ distribution with a skewed Gaussian (see the
\emph{solid curve} in Fig.~\ref{fig:ewvsd}) and compute the
probability of strong absorbers being less than 8 kpc to be $\sim
11\%$. The detection probability further increases to $\sim$25\% if we
consider the systems with \ew\ $\ge 2$~\AA. Recall that, about 95\% of
strong \mgii systems in our sample do not have \oii\ nebular emission
detected in SDSS fibres above the detection threshold. Therefore, the
above larger probability of strong \mgii absorber being at smaller
$\rho$ ($ \le 8 \rm ~kpc$) than the $\sim 1-3$\% detection rate of
nebular emission lines from \mgii absorbers (see,
Fig.\ref{fig:fraction}) either related to large intrinsic spread in
$\rho$ at a given \ew\ or a large spread in \loii\ at a given \ew.

\section{Conclusions}

Using the \mgii absorbers found in SDSS-DR7 and SDSS-DR12 spectra we
have complied a sample of $\sim 198$ \mgii systems with detectable
nebular emission lines over a redshift range of $0.36 \le$ \zabs\ $\le
1.1$. By studying the absorption and emission line properties of this
unique set of \mgii systems we derive the following results :

(1) The \mgii absorbers in our sample are found to be mostly sub-\ls
with luminosities ranging from $0.14-3.5$~\lsoii\ and
$0.09-3.5$~\lsoii, with a median of 0.54 and 0.43~\lsoii\ in the
SDSS-DR7 and SDSS-DR12, respectively. A typical SFR of the \mgii
systems uncorrected for dust reddening and fibre losses are found to
be in the range of $0.5-20 \rm ~M_{\odot} ~yr^{-1}$. We show that our
data suffers from finite fibre size effects and all the above quoted
values should be considered as lower limits.

(2) The nebular emission is preferentially detected in the strong
\mgii systems, i.e, $\sim$ 96\% of systems are having \ew\ $\ge
1$~\AA, with a mean \ew\ of $\sim$ 2.3~\AA. The detection rate of
nebular emission from \mgii absorbers with \ew $> 2$ \AA\ and
\loii\ $\ge $0.3\ls is found to be very small at 3\% and 1\% in the
SDSS-DR7 and DR12 spectra, respectively. We find the detection
probability to depend  on the \ew, where larger \ew\ systems
show higher detection rate. Furthermore, the detection rate increases
with the size of the fibre aperture (see Figure~\ref{fig:fraction}).
For a given \ew, the detection rate of nebular emission increases with
increasing $z$.

(3) In contrast to the strong correlation seen between \ew\ and
\sloii\ obtained from the stacked spectra, our \mgii systems with
nebular emission do not show any statistically significant correlation
between \sloii\ and \ew\ even when we restrict ourselves to narrow $z$
ranges. We conclude that the correlation observed in the stacked
spectra are dominated by the increase in nebular line detection
probability with \ew. In addition, we find that \sloii\ is strongly
correlated with redshift. This is also dominated by the fibre size
effect where at higher redshifts the projected size of the fibres will
cover a larger fraction of galaxy area than for galaxies at low
redshifts.

(4) We have found that the physical conditions in galaxies in our
sample evolve with redshift. A rise of $\sim 0.2$ dex in the nebular
emission line ratio of \o3o2 and \ohb\ is seen over a redshift range
of $0.1 \le z \le 1$. This is similar to what is seen in normal
galaxies at this redshift range. From the well known stellar mass
dependence on the line ratio we suggest that the \mgii absorbers
likely belongs to the population of low stellar mass galaxies with a
typical stellar masses in the range $9.2 < \rm log(M/M{_{\odot}}) <
9.4$.

(5) The typical velocity widths (\sigoii) of the \oii\ nebular
emission in \mgii absorbers are found in the range of $6-198$ \kms,
with an average \sigoii $\sim 75$ \kms. The \sigoii\ does not show any
correlation with $z$ and \ew. We measure the maximum relative velocity
shift ($|\Delta v|$) between emission and absorption to be $\sim
70$~\kms. By comparing the \sigoii\ and $|\Delta v|$ we show that most
systems are well within the regime where gas is bound to the host
galaxy. In addition, the \mgii absorbers probe the galaxies with range
of dynamical mass, i.e., log \mdyn ($M_{\odot}$) = 7.5-10.8.

(6) We show that \mgii systems follow the well known relationships
between \ew\ and impact parameter even at small impact parameters. A
comparison of their \ew\ distribution with that of the Milky Way shows
that the GOTOQs are most likely produced in the ISM+halo of their host
galaxy. Hence, these systems are ideally suited for probing various
feedback processes at play in $z < 1$ galaxies. In addition, by
comparing the doublet ratio ($DR$) and $R$ parameter of strong \mgii
absorbers with and without emission line detection we find that most
of the GOTOQs have large $R$ values and $DR \sim 1$. Therefore, good
fraction of these systems could be DLAs.

\section*{Acknowledgments}
 
We thank the anonymous referee for constructive comments and
suggestions. RS, PN, and PPJ acknowledge the support from Indo-French
Centre for the Promotion of Advance Research (IFCPAR) under project
number 5504$-$2.\par

Funding for the Sloan Digital Sky Survey IV has been provided by
the Alfred P. Sloan Foundation, the U.S. Department of Energy Office of
Science, and the Participating Institutions. SDSS-IV acknowledges
support and resources from the Center for High-Performance Computing at
the University of Utah. The SDSS web site is www.sdss.org.

SDSS-IV is managed by the Astrophysical Research Consortium for the
Participating Institutions of the SDSS Collaboration including the
Brazilian Participation Group, the Carnegie Institution for Science,
Carnegie Mellon University, the Chilean Participation Group, the
French Participation Group, Harvard-Smithsonian Center for
Astrophysics, Instituto de Astrof\'isica de Canarias, The Johns
Hopkins University, Kavli Institute for the Physics and Mathematics of
the Universe (IPMU) / University of Tokyo, Lawrence Berkeley National
Laboratory, Leibniz Institut f\"ur Astrophysik Potsdam (AIP),
Max-Planck-Institut f\"ur Astronomie (MPIA Heidelberg),
Max-Planck-Institut f\"ur Astrophysik (MPA Garching),
Max-Planck-Institut f\"ur Extraterrestrische Physik (MPE), National
Astronomical Observatories of China, New Mexico State University, New
York University, University of Notre Dame, Observat\'ario Nacional /
MCTI, The Ohio State University, Pennsylvania State University,
Shanghai Astronomical Observatory, United Kingdom Participation Group,
Universidad Nacional Aut\'onoma de M\'exico, University of Arizona,
University of Colorado Boulder, University of Oxford, University of
Portsmouth, University of Utah, University of Virginia, University of
Washington, University of Wisconsin, Vanderbilt University, and Yale
University.

\bibliography{references}

\label{lastpage}

\end{document}